\documentclass[%prl,
aps,%preprint %tightenlines
 ,twocolumn
]{revtex4}
\newcommand{\bea}{\begin{eqnarray}}

\newcommand{\beq}{\begin{equation}}

\newcommand{\bef}{\begin{figure}[btp]}

\newcommand{\bra}[1]{\langle #1 |}

\newcommand{\eeq}{\end{equation}}

\newcommand{\eea}{\end{eqnarray}}

\newcommand{\eef}{\end{figure}}

\newcommand{\exptn}[1]{\langle #1 \rangle}

\newcommand{\ket}[1]{| #1 \rangle}

\newcommand{\lket}[1]{| {\bf #1} \rangle}

\newcommand{\nnum}{\nonumber \\}

\newcommand{\op}[1]{{\hat{#1}}}

\usepackage{graphicx}

\begin{document}
%\draft
\title{Quantum Optical Systems for the Implementation of Quantum 
Information Processing}
\author{T.~C.~Ralph,\\
Centre for Quantum Computer Technology, Department of Physics,\\
University of Queensland, St Lucia 4072, Australia}
\date{\today}

\begin{abstract}
We review the field of Quantum Optical Information from elementary considerations through to quantum computation schemes. We illustrate our discussion with descriptions of  experimental demonstrations of key communication and processing tasks from the last decade and also look forward to the key results likely in the next decade. We examine both discrete (single photon) type processing as well as those which employ continuous variable manipulations. The mathematical formalism is kept to the minimum needed to understand the key theoretical and experimental results.
\end{abstract}

\maketitle

\vspace{10 mm}

\section{Introduction}

Information is not independent of the physical laws that govern how it is stored and processed \cite{LAN91}. The unique properties of quantum mechanics imply radically different ways of communicating and processing information \cite{NIE00}. However, to realize the potential of {\it quantum information} science, quantum systems with very special properties are needed. For example, it is essential that the quantum system evolves coherently and thus must be well isolated from the surrounding environment, but, in order that the information stored in the system can be processed and read out, it must also be possible to produce very strong interactions between the system and classical meters and control elements. In this paper we will review progress in achieving quantum information processing in optics, where the system in question is the quantum state of an electro-magnetic field mode at optical frequencies.

\subsection{Quantum Information}

It was perhaps Wiesner \cite{WIE72} who first realized that there are information tasks that can be achieved more effectively using quantum systems as the information carriers when he introduced his {\it quantum money} in 1972. The idea was to provide security against counterfeiting by encoding part of the bank note's serial number on quantum systems. This idea was famously extended to communications by Bennett and Brassard in 1984 \cite{BEN84} when they introduced {\it quantum key distribution}, a system for securely distributing a cryptographic key. Both of these applications depend on the unique property that quantum information can not be {\it cloned} \cite{WOO82}. That is, given a quantum system in an unknown state, it is not possible to produce an identical copy of the system (whilst retaining the original).

Other communication tasks that could be achieved  only with quantum systems started appearing in the early 1990's.  A key realization was that entanglement could be used as a resource for such tasks. A pair of spatially separated quantum systems are said to {\it entangled} if the state that describes the joint system cannot be factored into a product of states describing the individual systems. For example if two distant parties share entanglement then they can communicate classical information at twice the classical rate through the technique of quantum {\it dense coding} \cite{BEN92}. Similarly,  in the presence of entanglement, quantum information can be communicated via the exchange of classical information through the technique of quantum {\it teleportation} \cite{BEN93}. 

Around the same time that Bennett and Brassard were first describing quantum communication, Feynman \cite{FEY86} noted the possibility that computer algorithms existed that could be more efficiently processed by quantum systems than classical systems. Although toy examples of such algorithms were suggested by Deutsch soon after \cite{DEU86} it was not till 1995 that Shor \cite{SHO94} showed that an important problem, the determination of prime factors, could be solved in exponentially less time using a processor based on quantum systems, a {\it quantum computer}. The key technique, {\it quantum error correction}, was developed soon after \cite{SHO95,STE96}. This enables coherent correction of the logical errors which will inevitably creep into any calculation on a quantum computer. Another influential algorithm, showing speed up for the searching of an unsorted data base, was subsequently developed by Grover \cite{GRO97}. These developments showed that {\it fault tolerant} quantum computers (i.e. where errors can be corrected in the presence of imperfect gate operations) were in principle possible and that such machines could solve interesting problems. This in turn led to an explosion of interest in the field of quantum information.

Quantum information was originally discussed in terms of binary systems. Consider a two-level quantum system. This could be the spin states of an electron: up or down; two well isolated energy levels of an atomic system or; many other possibilities including various optical field states as we shall describe later. It is clear
that such two level systems could be used to carry {\it bits} of information.
For example, we could assign the value ``zero'' to one of the states, writing it in Dirac notation \cite{SAK85} as $|\bf 0 \rangle$,
and ``one'' to the other state writing $|\bf 1 \rangle $. A string of
these objects could then faithfully represent an
arbitrary bit string.

However, quantum objects offer more possible
manipulations than classical carriers of bits. In particular not only
can we have zero's and one's, but we can also have superpositions of
zeros and ones such as the plus state $|+ \rangle =
(1/\sqrt{2})(|\bf 0 \rangle + |\bf 1 \rangle)$. Indeed bits can just as
effectively be encoded in these superposition basis states, for example
using $|+ \rangle$ as a zero and $|- \rangle =
(1/\sqrt{2})(|\bf 0 \rangle - |\bf 1 \rangle)$ as a one. Because of these extra
degrees of freedom we refer to information digitally encoded on quantum
systems as quantum bits or $\it qubits$ \cite{SCH95}.

One non-classical feature of encoding in this way is the fact that
different bases do not in general commute.
Thus simultaneous, ideal measurements in
both bases cannot be made. Furthermore any measurements which obtain
information about the bit values of one basis
inevitably disturbs the bit values of the other basis. As we have mentioned these features (and more generally the no-cloning theorem) can be
used to create a secure communication channel via the technique of
quantum key distribution (also referred to as quantum
cryptography). 
%A number of demonstrations of quantum key distribution
%have been made in optics. 

Another feature of qubits is their ability to span all different bit
values simultaneously. This is obviously true of a single qubit where
the $|+ \rangle$ state, when viewed in the computational basis, $|\bf 0 \rangle$ and $ |\bf 1 \rangle$,
equally spans the two different bit values, 0 and 1.
This continues to be true for multi-qubit states. For example suppose we
start with two qubits in the state
\beq
|\bf 0 \rangle |\bf 0 \rangle
\eeq
where the first ket represents the first qubit and the second ket the second qubit and a tensor product is implied between their two Hilbert spaces. If we rotate both of them into their plus states we end up with
the state
\beq
|\bf 0 \rangle |\bf 0 \rangle + |\bf 0 \rangle |\bf 1 \rangle + |\bf 1  \rangle |\bf 0
\rangle + |\bf 1 \rangle |\bf 1 \rangle
\label{super}
\eeq
which is an equal superposition of all four possible two bit values.
This generalizes to $n$ qubits where the same operation of rotating
every individual qubit leads to an equal superposition of all
$2^{n}$ bit values.

Although this ability to span all possible inputs simultaneously hints
at the possibility of increased communication or computation power
using qubits, it is not the whole story. Note in particular that
analogues of the
sort of superpositions represented by Eq.~\ref{super} can also be
created in classical optical systems as superpositions of classical
waves. In order to unlock the full power of quantum information we
need to create entangled states such as the state
\beq
|\bf 0 \rangle |\bf 0 \rangle + |\bf 1 \rangle |\bf 1 \rangle
\label{ent}
\eeq
which clearly cannot be factored into contributions from the individual qubits and has no classical wave analogue.

If we consider information processing using qubits instead
of classical bits we need to introduce {\it quantum
gates}. Some of these will have classical counterparts, for example
the NOT gate takes $|\bf 0 \rangle$ to $|\bf 1 \rangle$ and vice versa. On the
other hand some gates will have no classical analogue, such as the
Hadamard gate which takes $|\bf 0 \rangle$ to $(1/\sqrt{2})(|\bf 0 \rangle + |\bf 1 \rangle)$
and $|\bf 1 \rangle$ to $(1/\sqrt{2})(|\bf 0 \rangle - |\bf 1 \rangle)$. We also require two qubit gates such as
the control-NOT (CNOT) \index{CNOT}
which preforms the NOT operation on one qubit
(the target) only if the other qubit (the control) has ``one'' as its
logical value. Eventually, if large arrays
of gate operations can be implemented efficiently, and fault tolerantly, on many qubits, one could consider
performing quantum computation. 
Although considerable progress has been made, the realization of
quantum computation experimentally still remains a long way off.

In more recent years quantum information research has been extended to systems with Hilbert space dimensions greater than two. In particular, there has been considerable interest in infinite dimensional Hilbert spaces and the quantum information properties of continuous degrees of freedom such as position and momentum \cite{BRA03}. Continuous variable versions of teleportation \cite{BRA98} and key distribution \cite{RAL00,HIL00} were developed early on and many other protocols followed. Quantum computation proposals based on continuous variables have also been developed \cite{GOT01,RAL03}.

\subsection{Quantum Optics}

The invention of the laser in the early sixties and its subsequent development led to an unprecedented increase in the precision with which light could be produced and controlled, and hence enabled the ability to systematically investigate the quantum properties of optical fields; {\it quantum optics}. The fundamental theoretical description of the quantized electromagnetic field was due to Dirac in the early days of quantum mechanics \cite{DIR58}. Stimulated by the new technological possibilities, Glauber \cite{GLA62}, Louiselle \cite{LOU73} and others laid the theoretical basis for the description of the laser and identified the signatures of non-classical light.

It was soon realized that quantum optics offered a unique opportunity to test fundamentals of quantum theory not previously available for experiments. The first experiments to demonstrate in a semi-controlled way the production of single light quanta or {\it photons} were arguably those of Kimble et al \cite{KIM77} based on the resonance fluorecense of single emitters, as proposed by Carmichael and Walls \cite{CAR76}. Pairs of photons produced by atomic cascades were shown to be in entangled states by Aspect, Grangier and Rogers \cite{ASP81} with non-classical correlations sufficiently strong to exclude all local-realistic hidden variable theories through violation of Bell inequalities \cite{BEL71}. These results followed from the earlier work of Clauser, et al \cite{CLA69} and Freedman \cite{FRE72} that adapted the original inequalities to the experimental setting. It should be noted that even now experimental efficiencies are not high enough to avoid the need for a fair sampling assumption in the data analysis of these types of experiments, thus not  closing all "loopholes" for these inequalities. Heralding of single photon states using atomic cascades \cite{GRA86} and {\it parametric down conversion} \cite{GHO87} followed. The latter technique uses a second order non-linearity to produce pairs of photons at half the pump frequency spontaneously, and has been the workhorse of photon experiments for the last twenty years. That the pairs of photons from down conversion can be made indistinguishable and hence exhibit Bosonic interference effects was shown in key experiments by Hong, Ou and Mandel \cite{HON87}.

{\it Squeezed states}, that exhibit non-classical statistics for their quadrature amplitudes, which are continuous variables, were discussed in the seventies \cite{YUE76} and eighties \cite{WAL83} and eventually demonstrated by Wu et al \cite{WU87}. Demonstration of entanglement between quadrature amplitudes, strong enough to demonstrate the paradox of Einstein, Podolsky and Rosen \cite{EIN35}, followed by Ou et al in the early nineties  \cite{OU92}.

Light can be described quantum mechanically in terms of the mode {\it annihilation operator} $\op a$, its conjugate, the {\it creation operator} $\op a^\dagger$ and the electromagnetic field mode ground, or {\it vacuum state} $\ket{0}$. The action of the creation operator on the vacuum state is to create a single photon {\it number state}, in a single spatio-temporal mode, i.e. $\op a^\dagger \ket{0} = \ket{1}$. In general $\op a^\dagger \ket{n} = \sqrt{n+1}\ket{n+1}$ where $n$ is a positive integer. Similarly the annihilation operator annihilates a single photon in a particular single spatio-temporal mode and in general  $\op a^\dagger \ket{n} = \sqrt{n}\ket{n-1}$. The number states form an ortho-normal basis convenient for representing arbitrary states. The mode operators obey the commutation relation $[\op a, \op a^\dagger] = 1$. It is often convenient to pick our mode decomposition in terms of single frequency eigenstates using the nomenclature $\op a_\omega$ and $\ket{0}_\omega$. Then we have
\beq
[\op a_{\omega'}, \op a_\omega^\dagger] = \bra{0}_{\omega'}\ket{0}_\omega = \delta(\omega'-\omega)
\label{com}
\eeq

The optical observables of interest are the photon number, $\op n = \op a^\dagger \op a$, and the quadrature amplitude, $\op X^\theta = e^{i \theta} \op a + e^{-i \theta}  \op a^\dagger$. Photon number is proportional to intensity for bright fields and can be measured by photo-detectors. For dim fields individual photons can be resolved with photon counters. The quadrature amplitude of the field can be measured by beating the signal field with a bright, phase reference field at the same optical frequency, a {\it local oscillator} (LO), and then measuring it with by photo-detection. This is known as {\it homodyne} detection. The angle $\theta$ is the phase difference between the signal and the LO and is usually taken to be in-phase ($\theta=0$) or out-of-phase ($\theta=\pi$), giving two conjugate (i.e. non-commuting) variables analogous to position and momentum.

As well as the number states, another key state in quantum optics is the coherent states. The coherent states are displaced vacuum states defined by 
\beq
\ket{\alpha} = \op D(\alpha) \ket{0}
\eeq
where the displacement operator is
\beq
\op D(\alpha) = e^{i(\op a \alpha + \op a^\dagger \alpha^\ast)}
\eeq
The coherent states are eigenstates of $\op a$ with eigenvalue $\alpha$. This leads to average values for their quadrature observables that are the same as for a classical field with the same amplitude. Hence the coherent state is often thought of as the quantum mechanical state which is the closest approximation to a classical optical field. The output of a well stabilized laser is a mixed state which can be approximately decomposed as an ensemble of coherent states with fixed magnitude but random phases \cite{MOL97}. However, in situations where the phase is unimportant, or when the LO is derived from the same laser as the signal such that the phase is common mode, it is convenient to model laser output as being in a single coherent state of fixed magnitude and phase.

Because of the success in demonstrating fundamental quantum effects in optics, light was an obvious candidate for demonstrating the predictions of quantum information science. Here we will review quantum optics successes in quantum information science and look at its potential for achieving more complex quantum processing tasks in the future. 
%as the only reasonable candidate for quantum communication. This is not just because of its mobility, but also because of the ease with which certain critical manipulations of quantum optical states can be achieved. 

%We discussed in section (\ref{sec:OPTICALCOMMUNICATION}) communicating
%information using light. We saw that the quantum nature of light
%leads to intrinsic limits on the amount of information that can be
%sent optically. We looked at experiments using squeezed light which
%could reduce those limits as much as possible. In those discussions we
%imagined encoding the information on the light
%in much the same way as one would if
%sending information via a classical medium.

%We now consider a quite different idea: that of encoding
%information in ways only possible for quantum systems. Information
%encoded in this way is referred to as {\it Quantum Information}. From
%this point of view we will find communication and computation tasks that,
%rather than being limited by quantum mechanics, are actually enhanced
%by it.

%The technical challenges in working with quantum information are
%formidable. Never-the-less significant experimental demonstrations
%have been made, a number of which we will discuss here. As we shall
%see quantum optics has some considerable advantages (and disadvantages)
%over alternative quantum systems from a quantum information point of
%view.

\section{Encoding Classical Information on Light}
\label{CIL}

Before considering the quantum information potential of optics we first discuss the encoding of classical information on quantum states of light. Current optical communications systems operate in a regime in which quantum effects can be ignored. In the future, as higher and higher communication efficiency is required, this is likely to change. Here we consider the ultimate limits imposed by quantum mechanics.
We quantify this using the {\it channel
capacity}, a concept that describes the maximum amount of information that can
be transmitted based on statistical arguments. More detailed reviews of the techniques for
the encoding, propagating and decoding of information on quantum
systems can be found in Ref.~\cite{YAM86} and Ref.~\cite{CAV94}.

The Shannon capacity \cite{SHA48} of a communication channel
operating at the bandwidth limit is
\begin{equation}
C={{1}\over{2}} \log_{2}[1+{{S}\over{N}}]
\label{cc}
\end{equation}
where $N$ is the noise power (variance), assumed Gaussian, and $S$ is the
signal power, also assumed Gaussian distributed. Here $C$ is in units of bits per symbol. Eq.\ref{cc} 
can be used to calculate the channel capacities of quantum
states with Gaussian probability distributions such as coherent states
and squeezed states. Consider first a signal composed of a
Gaussian distribution of coherent state amplitudes all displaced at the
same quadrature angle, say $\theta = 0$ ($\alpha$ real).
The signal power $V_{s}$ is given by the variance
of the distribution. The noise is given by the intrinsic quantum
noise of the coherent states, $V_{n}=\exptn{\op X^2}-\exptn{\op X}^2 =1$.
Because the quadrature angle of the signal is known, homodyne
detection can in principle detect the the signal without further
penalty, thus the measured signal to noise is $S/N=V_{s}/V_{n}=V_{s}$.

In general the average photon number per bandwidth per second
of a light beam is given
by
\begin{equation}
\bar n = {{1}\over{4}}(V^{+}+V^{-})-{{1}\over{2}}
\label{n}
\end{equation}
where $V^{+}$ ($V^{-}$) are the variances of the maximum (minimum)
quadrature projections of the noise ellipse of the state. These
projections are orthogonal quadratures, such as in-phase and out-of-phase,
and obey the uncertainty principle $V^{+}V^{-} \ge 1$. In the above
example one quadrature is made up of signal plus quantum
noise such that $V^{+}=V_{s}+1$ whilst the orthogonal quadrature is just
quantum noise so $V^{-}=1$. Hence $\bar n= 1/4 V_{s}$ and so the
channel capacity of coherent states with single quadrature encoding
and homodyne detection is
\begin{equation}
C_{c}=\log_{2}[\sqrt{1+4 \bar n}]
\label{cccho}
\end{equation}
Showing in an experiment that a particular optical mode has this
capacity would involve: (i) measuring the quadrature
amplitude variances of the beam, $V^{+}$ and $V^{-}$,
(ii) calibrating the sender's signal
variance and (iii) measuring the receiver's signal to noise. If these
measurments agreed with the theoretical conditions above then
Shannons theorem tells us that an encoding scheme exists which could
realize the channel capacity of Eq.\ref{cccho}. An example of such an
encoding is given in \cite{CER01}.

If the average photon number per symbol is such that $\bar n > 2$, improved channel capacity can be
obtained by encoding symmetrically on orthogonal quadratures and detecting both
quadratures simultaneously using a 50:50 beamsplitter followed by dual homodyne
detectors, one for each quadrature (equivalently heterodyne detection can be used). Because of
the non-commutation of orthogonal quaratures there is a penalty for
their simultaneous detection which reduces the signal to noise of each
quadrature to $S/N=1/2 V_{s}$. Also because there is signal on
both quadratures the average photon number of the beam is
now $\bar n= 1/2 V_{s}$. On the other hand the total
channel capacity will now be the sum of the two independent channels
carried
by the two quadratures. Thus the channel capacity for a coherent
state with dual quadrature encoding and heterodyne detection is
\begin{eqnarray}
 C_{ch} & = & {{1}\over{2}}
\log_{2}[1+{{S}\over{N}}^{+}]+{{1}\over{2}}
\log_{2}[1+{{S}\over{N}}^{-}] \nonumber\\
& = & \log_{2}[1+\bar n]
\label{ccchet}
\end{eqnarray}
which exceeds that of the single quadrature homodyne technique (Eq.\ref{cccho}) for $\bar n>2$.

If we restrict
ourselves to a semi-classical treatment of light the above channel capacities are the best achievable. However the  channel
capacity
of the homodyne technique can be improved by the use of
non-classical,
squeezed light \cite{YUE78}. With squeezed light the noise variance of the encoded
quadrature
can be reduced such that $V_{ne}<1$, whilst the noise of the unencoded
quadrature is increased such that $V_{nu} \ge 1/V_{ne}$. As a result
the signal to noise is improved to $S/N=V_{s}/V_{ne}$ whilst the
photon number is now given by Eq.\ref{n} but with
$V^{+}=V_{s}+V_{ne}$ and $V^{-}=1/V_{ne}$, where a pure (i.e.
minimum uncertainty) squeezed state has been assumed. Maximizing the signal
to noise for fixed $\bar n$ leads to $S/N=4(\bar n+\bar n^{2})$ for a
squeezed quadrature variance of $V_{ne,opt}=1/(1+2 \bar n)$.
Hence the channel capacity for a squeezed beam with homodyne
detection is
\begin{equation}
C_{sh}=\log_{2}[1+2 \bar n]
\label{ccsho}
\end{equation}
which exceeds both coherent homodyne and heterodyne for all values
of $\bar n$.
%
%Although squeezing gives an improvement in channel capacity this 
%improvement is rapidly washed out by non-idealities in the system such 
%as propagation loss and other inefficiencies. It still seems some way 
%off before squeezing might be considered a viable option for 
%increasing channel capacity.

In principle, a final improvement in channel capacity can be obtained by allowing
non-Gaussian states. The absolute maximum channel capacity for a
single mode is given by the Holevo bound and can be realized by
encoding in a maximum entropy ensemble of number states and
using photon number detection.
This ultimate channel capacity is
\begin{equation}
C_{Fock}=(1+ \bar n) \log_{2} [(1+\bar n)]-\bar n \log_{2} [\bar n]
\label{ccf}
\end{equation}
which is the maximal channel capacity at all values of $\bar n$.

\section{Optical Qubits}
\label{sec:OQ}

We now consider how quantum information can be carried by light. There are a number of ways in which qubits can be encoded optically which fall into two broad classes: dual rail and single rail encoding. In dual rail encodings two orthogonal quantum optical modes are used. In single rail encoding only one quantum optical mode is used, although a second classical mode is implicitly needed as a phase reference. In the following we will begin by describing these encoding techniques and discussing a number of examples. We will then focus on dual-rail encoding and discuss current experimental approaches and future prospects for "better" qubits.

\subsection{Dual Rail Encoding}
\label{subsec:DRE}

Consider two orthogonal optical modes represented by the annihilation operators $\op{a}$ and $\op{b}$ and the vacuum modes $\ket{0}_a$ and $\ket{0}_b$. For brevity we will write $\ket{0}_a \otimes \ket{0}_b \equiv \ket{00}$. We define our logical qubits as $\lket{0} = \op{a}^\dagger \ket{00} = \ket{10}$ and $\lket{1} = \op{b}^\dagger \ket{00} = \ket{01}$. That is single photon occupation of one mode represents a logical zero, whilst single photon occupation of the other represents a logical one. This is dual rail encoding.

\begin{figure}[htb]
\begin{center}
\includegraphics*[width=6cm]{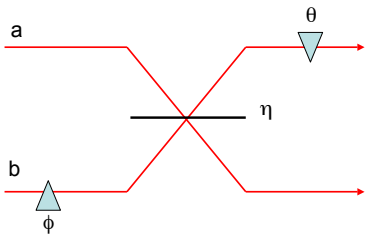}
\caption{Beamsplitter and phase-shifter circuit for producing an arbitrary single qubit evolution on a spatial dual-rail qubit }
\label{fig1}
\end{center}
\end{figure}

For example suppose $\ket{0}_a$ and $\ket{0}_b$ are spatio-temporal modes with identical profiles, polarization and centre frequency, synchronized in time, but spatially separated. Arbitrary single qubit operations can be achieved using a beamsplitter and two phase shifters as illustrated in Fig.~\ref{fig1}. A beamsplitter is a partially reflecting mirror that can coherently combine two optical modes in a set ratio. The interaction in the figure produces the following Heisenberg evolution of the mode operators:
\begin{eqnarray}
\op{a} & \to & \sqrt{\eta} \op{a} +e^{i \theta} \sqrt{1-\eta} \op{b} \nonumber\\
\op{b} & \to & e^{i \phi}(\sqrt{1-\eta} \op{a} -e^{i \theta} \sqrt{1-\eta} \op{b})
\label{hbs}
\end{eqnarray}
where $\eta$ is the intensity reflectivity of the beamsplitter. We have assumed the optical elements are lossless, a reasonable assumption for modern components. We have also assumed perfect mode matching between the two input modes to the beamsplitter, something rather more difficult to arrange in practice. Eq.\ref{hbs} implies the following qubit evolution \cite{BAC04}:
\begin{eqnarray}
\ket{10} & \to & \sqrt{\eta} \ket{10} +e^{i \phi} \sqrt{1-\eta} \ket{01} \nonumber\\
\ket{01} & \to & e^{i \theta}(\sqrt{1-\eta} \ket{10} -e^{i \phi} \sqrt{1-\eta} \ket{01})
\label{sbs}
\end{eqnarray}
which corresponds to an arbitrary single qubit unitary. 

\begin{figure}[htb]
\begin{center}
\includegraphics*[width=6cm]{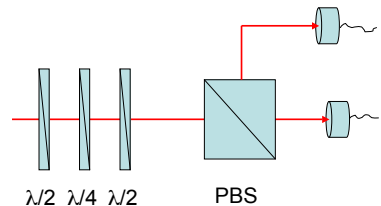}
\caption{Combination of half and quarter wave plates oriented at particular angles, a polarizing beamsplitter (PBS) and photon counting, enables detection of polarization dual-rail qubits in any basis. }
\label{fig2}
\end{center}
\end{figure}

More commonly two identical spatio-temporal modes but with different polarizations, say horizontal and vertical, will be used as the dual rails. Then we may write $\lket{0} = \ket{10} = \ket{H}$ and $\lket{ 1} = \ket{01} = \ket{V}$. Half and quarter-wave plates replace the phase shifters and beamsplitters in achieving arbitrary unitaries \cite{DOD03}. In particular the key operation the {\it Hadamard} gate, defined by $\lket{0} \to \lket{0} + \lket{1}$ and $\lket{1} \to \lket{0} - \lket{1}$, is implemented by a half wave-plate oriented at 22.5 degrees to the optic axis. Detection in any basis can be achieved via wave plates and polarizing beamsplitters which effectively converts polarization encoding into spatial encoding (see Fig.~\ref{fig2}). The ease of manipulation and stabilty of polarization states has made this encoding the most popular in optics.

Another possibility is a temporal encoding in which the dual rails are spatio-temporal modes which are identical except for a time displacement \cite{STU02}. These can again be manipulated at the single qubit level with linear optics, though not deterministically unless fast electro-optic switches are available.

A final possibility is a frequency encoding in which, this time, the dual rail modes are identical except for a frequency off-set. Here an active element is required in order to move power coherently between different frequencies. If the frequency off-set is in the radio to micro-wave band then acousto-optic modulators and asymmetric interferometers can be used for this purpose \cite{HUN04}. Although the most difficult to manipulate the frequency encoding would likely be the most robust to fibre transmission.

\subsection{Single-Rail Encoding}
\label{subsec:SRE}

Single-rail encoding requires only a single quantum mode, that can be placed into the states $\lket{0} = \ket{\phi}$ and $\lket{1} = \ket{\psi}$ or any superposition of them. The only requirement on these states is that they are orthogonal, i.e. that $\langle \phi \ket{\psi} = 0$. In general such qubits will be non-stationary, so a good "clock" is required in order to detect and manipulate them. In optics this clock, or phase reference, will typically be a classical optical mode derived from the master laser which is driving all the optical modes, in other words a local oscillator (LO). 

Perhaps the simplest choice for $\ket{\phi}$ and $\ket{\psi}$ are the vacuum and single photon states, such that $\lket{0} = \ket{0}$ and $\lket{1} = \ket{1}$. Producing and manipulating superposition states of the form $\mu \ket{0} + \nu \ket{1}$, as required for this type of qubit is not so easy. However, a universal set of non-deterministic operations have been described \cite{LUN02} and superposition states have been produced non-deterministically in experiments \cite{LVO02,BAB04}. One important feature of the single-rail encoding is that it is relatively easy to produce entangled states. If a single photon is split on a 50:50 beamsplitter the resulting state is $(1/\sqrt{2})(\ket{0} \ket{1} + \ket{1} \ket{0})$ which is a maximally entangled two qubit state in the single-rail encoding. Such states can then be used as a resource for quantum processing tasks. 

Another possible choice for $\ket{\phi}$ and $\ket{\psi}$ are two different coherent states, such that $\lket{0} = \ket{\alpha}$ and $\lket{1} = \ket{\beta}$. In general such states will not be orthogonal but their overlap is given by $|\langle \alpha \ket{\beta}|^2 = \exp[-|\alpha-\beta|^2]$ which is very small for quite modest differences in the amplitudes of the coherent states. A popular choice is to take $\beta = -\alpha$. By choosing  $\alpha \ge 2$ a negligible overlap is achieved. The computational states, $\ket{\alpha}$ and $\ket{-\alpha}$ can be distinguished via homodyne detection. A useful feature of this choice  is that the equal superposition state $\ket{\alpha} + \ket{-\alpha}$ ($\ket{\alpha} - \ket{-\alpha}$) contains only even (odd) photon number terms and so these orthogonal diagonal states can be distinguished by photon counting. A number of groups have discussed quantum information tasks using this encoding \cite{COC98,ENK02,JEO02,RAL03}. As for the single photon single-rail scheme, single qubit unitaries are difficult with this encoding but entanglement production is relatively easy. Indeed, splitting a superposition state like $\ket{\alpha} + \ket{-\alpha}$ many times on a beamsplitter leads to multi-mode entanglement. Whilst production of the coherent computational states is straightforward, to date the only experimental realizations of the superposition states have come from cavity quantum electro-dynamics experiments \cite{BRU96,TUR95}, though promising schemes \cite{LUN04} and initial results \cite{WEG04} suggest small traveling wave superposition states may be possible in the near future.

More exotic optical states have also been suggested for single rail encoding \cite{GOT01} that have nice error correction properties, but these are likely to be more difficult again to produce experimentally.

\subsection{Postselection and Coincidence Counting}
\label{sec:PSANDCB}

Producing and detecting single photon states efficiently is a
major technological challenge. Currently the best single photon detectors have efficiencies around 90\% and the most efficient single photon sources are around 55\%, but typically in practical situations these numbers are much lower. This presents a major problem for single rail schemes where typically the loss of a photon results in a change to the qubit state and hence logical errors. In dual-rail schemes on the other hand, photon loss results in no qubit arriving (rather than the wrong qubit) and so can quite easily be filtered out of the data as we shall now describe. This is another reason why most optical quantum information demonstrations are currently based on dual-rail logic. 
%section (\ref{sec:SINGLEPHOTONDDET})
%we saw that current commercial photon counting detectors
%are ``off/on'' type detectors which
%only ``click'' when there are one or more photons incident in a
%particular time window. Typical maximum detection efficiencies are
%around 70\%.

%Light sources which produce very low photon numbers tend to be {\it
%spontaneous sources}, that is they produce photons randomly. In the
%next section we will examine experimental progress towards creating
%light sources which produce single photons on demand. Here we will
%discuss how in principle experimental demonstrations can be performed
%using on/off detectors and spontaneous sources.

We begin by discussing how single photon experiments can be performed by strongly attenuating a 
single mode laser source.
We can represent the state of such a laser source by the state
\beq
|\psi \rangle = |0 \rangle + \alpha |1 \rangle + {{\alpha^{2}}\over{2!}}
|2 \rangle +\ldots
\eeq
As we attenuate the source more and more, $\alpha$
becomes much less than one and we can write to a good approximation
\beq
|\psi \rangle = |0 \rangle + \alpha |1 \rangle
\eeq
We now have a source which in any particular time interval (length determined
by the frequency dependence) has a high probability of producing
vacuum; some small probability of producing a single photon state and;
a negligible probability of producing a multi-photon state. If a
photon counter is placed at the end of the experiment and we only
worry about those times when the detector ``clicks'' then we will
{\it postselect} just the single photon part of the state. If the
source is polarized then it can be manipulated as a dual-rail qubit. Notice,
however, that it is a rather inconvenient qubit source as it rarely
works and you only know it worked after the fact, by evaluating
the detection
record. Never-the-less this
type of source
has successfully been used to demonstrate single qubit type experiments
such as quantum key distribution (see section \ref{SEC:QKD}).

A major problem arises with an attenuated coherent source if we try to
move to experiments requiring two qubits. One might assume we could
use two highly attenuated coherent sources and then postselect only
those events where two photons appear at the end of the experiment.
However, the joint state of two equal power, attenuated lasers is
\bea
|\psi \rangle_{ab} & = & (|0 \rangle + \alpha |1 \rangle +
{{\alpha^{2}}\over{2!}} |2 \rangle +\ldots)_{a} \nonumber\\
& & \;\;\;\;\; (|0 \rangle +
\alpha |1 \rangle + {{\alpha^{2}}\over{2!}} |2 \rangle +\ldots)_{b} \nonumber\\
& = & |0 \rangle_{a} |0 \rangle_{b} + \alpha (|1 \rangle_{a} |0
\rangle_{b} +
|0 \rangle_{a} |1 \rangle_{b}) + \nonumber\\
& & \; \; \; {{\alpha^{2}}\over{2!}}(2! |1 \rangle_{a} |1
\rangle_{b} +
|2 \rangle_{a} |0 \rangle_{b} + |0 \rangle_{a} |2 \rangle_{b}) + \ldots
\label{EQ:2COH}
\eea
where the first ket refers to one source whilst the second one to the other
source. Notice that if we go to order $\alpha^{2}$ then there is
indeed a term with a single photon state in each beam. However, terms
involving pairs of photons in one beam with vacuum in the other occur
with the same probability. Postselecting on two photon events will not
in general remove these terms. Hence it is not possible in general to
perform two
qubit experiments using highly attenuated laser sources. A more
sophisticated solution is required.

Since
the late eighties, the solution of choice has been parametric
down conversion in a $\chi_2$ medium \cite{GHO87}.  Weak parametric down conversion results in the spontaneous converstion of single pump photons at the harmonic frequency into 
pairs of photons at the fundamental. If the down conversion is spatially non-degenerate
then, in the Schr\"odinger picture,
initial vacuum inputs are transformed according to
\beq
|0 \rangle_{a} |0 \rangle_{b} \to 
(|0 \rangle_{a} |0 \rangle_{b} + \chi' |1 \rangle_{a} |1 \rangle_{b}
+ \chi'^{2} |2 \rangle_{a} |2 \rangle_{b} + \ldots)
\eeq
where $\chi'$ is an effective non-linear interaction strength, proportional to the pump power. If we now allow $\chi'$ to be very small, which is not hard to arrange experimentally, then the 
state produced is given to an excellent approximation by
\beq
|\psi \rangle_{ab} =
|0 \rangle_{a} |0 \rangle_{b} + \chi' |1 \rangle_{a} |1 \rangle_{b}
\label{EQ:2PHOTON}
\eeq
In contrast to equ.(\ref{EQ:2COH}) the state in equ.(\ref{EQ:2PHOTON}) has
only the desired two photon term to first order in $\chi'$. If we
postselect only those events from the detection record in which 2
photons are detected ``simultaneously'' or in coincidence
(within some preset time window) then
we will only record the part of the state which is due to the pairs of photons.
Thus by using the combination of parametric down-conversion, the 
polarization degree of freedom
and postselection,  we can perfom, at least in principle,  2 qubit experiments. 
Experiments carried out this way are sometimes referred to as
coincidence basis experiments and we will discuss various examples in later sections.
However, note that this source is still spontaneous, ie
successful events are rare, random and
we do not know if they have occurred until after the fact. Although, 3 and 4 qubit experiments have been achieved by a simple generalization of the techniques just outlined, the cost is an exponential drop in the probability of success.
%, and so direct scaling to much higher qubit numbers is impractible. 
Thus, although experiments carried out in coincidence can demonstrate the basic physics of particular systems, they are intrinsically not scaleable to large scale quantum information processing.
Progress in
producing sources without this drawback are discussed in the next
section.

\subsection{True Single Photon Sources}
\label{sec:TSPS}

We now discuss two distinct approaches to producing better
approximations to single photon states. The first is to create a
{\it heralded} single photon source. That is a source which, though
not always producing a single photon state, produces a clear signal
when successful. Such a source could be made semi-deterministic by the use of quantum memory. 
The second approach is to produce an {\it on-demand}
source, which deterministically produces a single photon state when
requested.

\subsubsection{Heralded Single Photons}

By detecting one of the output modes and only accepting the other
output if a photon is detected, a heralded single photon source can be created using spatially
non-degenerate down-conversion. From an idealistic point of view the
conditional state when a single photon is detected in mode $a$ can be
obtained from equation (\ref{EQ:2PHOTON}) as
\beq
\bra{1}_{a}|\psi \rangle_{ab} =  \chi'  |1 \rangle_{b}
\label{EQ:2PHOTONC}
\eeq
indicating that a single photon state is created in mode $b$ with
probability $|\chi'|^{2}$. In reality things are not so simple.

We expand the output state of the down-converter in a basis of wave-number eigenstates, each defining a single frequency spatial mode, to obtain 
\beq
|\psi \rangle_{ab} =
|0 \rangle_{a} |0 \rangle_{b} + \chi' \int dk_{a} dk_{b} F(k_{a},k_{b})
|1 \rangle_{a} |1 \rangle_{b}
\label{EQ:2PHOTON2}
\eeq
where $k_{i}$ is the wave vector of the $i$th beam and the function
$F(k_{a},k_{b})$ describes the spatio-temporal structure of the
modes. The intrinsic spatio-temporal resolution of the photon counter far exceeds the read-out resolution. Thus the photon counter selects an ensemble of distinguishable
single photon modes of which the experimenter is ignorant. This situation can be described by the mixed state
%(see section (\ref{sec:MIXEDSTATES}))
%
\beq
\rho_{a} = \int dk_{a} T(k_{a}) \ket{1}_{a}\bra{1}_{a}
\eeq
where $T(k_{a})$ is the spatio-temporal distribution of the detected
ensemble. The output state, conditional on a photon count, is then
\beq
\rho_{b} = Tr_{a}[\ket{\psi}_{ab}\bra{\psi}_{ab} \rho_{a}]
\eeq
In general $\rho_{b}$ is a mixed state, however if $T(k_{a})$ is
centred on but much ``narrower'' (both spatially and temporally) than
$F(k_{a},k_{b})$, then to a good approximation the pure single photon
number state
\beq
\ket{\psi}_{b} = \chi' \int dk_{b} T(k_{b}) \ket{1}_{b}
\eeq
is produced. 

Perhaps the most conclusive demonstration of photon production by this (or indeed by any) method was the experiment by Lvovsky et al.\cite{LVO01}.   A beta-barium borate crystal was used in a type I arrangement to produce frequency degenerate but spatially non-degenerate photon
pairs. Transform limited pulses at 790[nm] from a Ti:Sapphire laser
were doubled and used to pump the crystal. Dispersion tends to make the pump and signal beams follow different paths through the crystal. Great care was taken to
minimize any distortions of the spatial and temporal modes of the
outputs due to this walk-off of the pump beam. The trigger photons were passed through a
spatial filter and a $0.3$ [nm] frequency filter before being counted
on a single photon detector. The success of the heralding process was characterised by
performing homodyne tomography on the conditionally produced photon
states. For best results the local oscillator pulse (LO) used in the homodyne detection
of the signal must be accurately mode-matched to the single photon
state. Mode matching with a visibility of about $80$\% was achieved.

The homodyne data collected was then used to produce the Wigner
distribution of the single photon state \cite{SMI93}. The {\it Wigner distribution} is a quasi-probability function for the quadrature amplitudes of the field state. The marginal distributions for the individual quadratures obtained from the Wigner function correspond to their respective probability distributions.
Normalization of the distribution was achieved by
simultaneously collecting vacuum state data. The resulting Wigner
function was consistent with a
mixed state comprised of $55$\% single photon state and $45$\%
vacuum. Although imperfect the single photon state was still pure enough to display negative regions in the Wigner distribution, a highly non-classical effect that demonstrates the strongly quantum mechanical nature of the detected single photon state.
%

%%
%%
%%Figure. Figure 4(c) from LVO01
%\bef
%\vspace{50mm}
%%\leavevmode \epsfbox{Ou92Fig1.eps}
%\caption{\it Reconstructed Wigner function from experiment of Lvovsky
%et al.\cite{LVO01} showing negative values close to the origin.}
%\label{fig:lvores}
%\eef

\begin{figure}[htb]
\begin{center}
\includegraphics*[width=10cm]{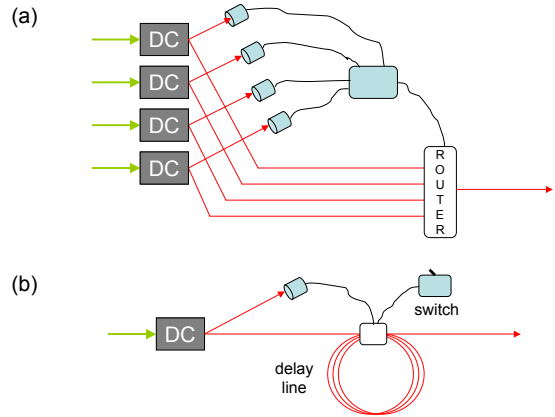}
\caption{Two schemes for producing pseudo on-demand sources from heralded sources. In (a) an array of downconverters (DC) are pumped simultaneously. Photon counters monitor one of the output modes of each DC to see if it fired. If one of the counters triggers the other mode of the corresponding DC is sent to the output via an optical router. For a sufficiently large array the probability of at least one DC producing a pair becomes large. In (b) a single DC is pumped. If the trigger detector fires a fast optical switch captures the photon in a delay line. At some later time the photon can be released on demand by flicking the optical switch open.}
\label{fig3}
\end{center}
\end{figure}

Lvovsky et al's results clearly show that a single photon state
can in principle be produced in this manner but it also highlights
problems with this approach. For example, in order to obtain
a conditional state which is as pure as possible, strong attenuation 
was applied to the
trigger photon resulting in low single photon state production rates of
about one photon every 4 seconds. Also,  mode matching of the single
photon state is seen to be a difficult problem. A more promising and practical
solution would be to mode-match the single photon state into an
optical fibre for subsequent use down-stream. The best results to
date for this difficult problem were achieved by
Kurtsiefer, Oberparleiter, and Weinfurter \cite{KUR01} who 
obtained about $40$\% single photon
contribution to the conditional state,

Finally, it is clearly inconvenient that the photons in these experiments
are produced at random times. A possible solution to this problem suggested by Migdall, Branning and Casteletto \cite{MIG02} is to pump many crystals simultaneously so that the probability that at least one of them produces a pair is high. Electro-optic switches could then route the successfully triggered photon into the output mode (see Fig.~\ref{fig3}(a)). Another solution to this problem, proposed
and demonstrated in principle by Pittman, Jacobs, and Franson
\cite{PIT02},
is to inject the single photon state into an optical fibre storage loop
when the trigger photon is detected. The captured photon is then held
there till required, when it is switched out of the loop (see Fig. \ref{fig3}(b)). If the
round trip time of the loop
is matched to the repetition rate of the pulsed pump laser then a
number of loops can be loaded and then released simultaneously to
produce several single photons states at the same time. Currently (as
well as the mode matching problem discussed above) the losses
associated with the Pockell cells used to switch the photons are prohibitively large. 

\subsubsection{Single Photons on Demand}

The dream of a push-button single photon source can most nearly be realized by
generating light from a single isolated emitter such as a single
ion or atom. The trick here is that a single emitter can only produce
a single photon ``at a time'' with some dead time between emissions
while the source is re-excited. The effect is that the output state
can be written (in an idealised fashion) as
\beq
|\psi \rangle = |0 \rangle + \alpha |1 \rangle + \tau
({{\alpha^{2}}\over{2!}} |2 \rangle +\ldots)
\label{TAU}
\eeq
where $\tau$ is a number between $0$ and $1$ representing the
suppression of higher photon number terms. If $\tau$ is very small
then $\alpha$ can be made large, such that there is a high
probability that a single photon will be emitted, whilst the probability
of multiple photon emission remains very low.

The first experiments of this kind were
performed in the late seventies \cite{KIM77}.
However, although they clearly displayed the photon anti-bunching
expected of a single photon source, they were very inefficient because
they radiated into $4 \pi$ steradians and, being based on atomic beams,
there was little that could be done to improve matters. More recently
various attempts have been made to
create more efficient single emitters. These included: placing single
neutral atoms or ions into high finesse optical cavities
\cite{KUH02,KEL04} such that the photon emission should be into a
single Gaussian mode; close coupling to single solid-state emitters
such as neutral vacancy (NV) centre in diamond \cite{BEV02} and;
the construction of single
quantum-dot emitters integrated into distributed Bragg reflector
(DBR) cavities \cite{SAN02}.

Initial experiments on quantum dot emitters were carried out by Santorini et al. \cite{SAN02}. Self-assembled
InAs quantum dots embedded in GaAs were sandwiched between
DBR mirrors to form tiny, high finesse,
monolithic cavities in the form of $5$ micron high pillars. These were
then cooled to 3-7 [K] and pumped by a pulsed Ti:Sapphire laser. The
quantum dot emission, at around 935[nm],
was spectrally filtered (0.1 [nm]) and a  single
polarization was selected before being coupled into single-mode
optical fibre. The efficiency of single photon production was
estimated to be about $30$\%.

The single-photon states thus produced were
characterized by their $g^{(2)}$ factor which was typically of the order of
$0.06$ ($\tau^{2} \approx g^{(2)}$) showing good suppression of two photon
emission. To test the indistinguishability of
the photon states the Hong-Ou-Mandel dip \cite{HON87}
between consecutive emissions was measured. The inferred
visibility of the dip
when measurement imperfections were taken into account was about $70$\%.

Intrinsic time-jitter due to the spontaneous excitation process employed has been identified as the prime cause of the loss of photon indistinguishability between pulses in the quantum dots \cite{KIR03}. 
The prospects for indistinguishability between independent emitters are more remote due to the inherent variability of the structures. A different approach which does not suffer from these drawbacks is to place a single ion in a high finesse cavity, as demonstrated by Keller et al \cite{KEL04}. A single calcium $40$ ion was trapped inside an 8mm, high finesse cavity with a $1.2$ MHz decay rate. After laser cooling, a photon is produced through a cavity assisted Raman process, which is coherent and does not suffer from inhomogeneous broadening. Suppression of two photon events was $\tau^{2} \approx 0.01$ and detection limited. The device could produce a stream of single photon events over more than an hour, at a repetition rate of $100$ KHz and a photon production efficiency of about $5$\%. Photon temporal indistinguishability was confirmed through observation of the photon wave-packet spread via the photon arrival time probability distribution.

The advantages in purity of the ion-trap photon source over the quantum dots comes at the price of a much more complicated set-up and much slower repetition rates. For both systems, the modest efficiencies mean that they are still effectively spontaneous sources. However, progress is rapid and we may anticipate systems combining the best aspects of the present systems with high efficiency in the not too distant future.

\subsection{Characterising photonic qubits and processes}
\label{sec:CHARPQ}

We now discuss the characterization of photonic qubits and processes. Our analysis assumes postselection, i.e. we only consider events in which a photon is detected. Of course this characterization
cannot be carried out for a single event because of the probabilistic
interpretation of quantum mechanics. But given a large ensemble of
detection events, corresponding to identically prepared photons and/or interactions, a
recipe can be given for determining the state of the ensemble or the process through which the ensemble was evolved.

We consider polarization encoded qubits. The polarization state of
the photons is most generally described by the density operator
$\op{\rho}$. Our observables are the
Stokes operators (corresponding to the classical Stokes parameters
\cite{STO52})
\bea
\op S_{1} & = & \op n_{H} - \op n_{V} = \ket{H}\bra{H} -
\ket{V}\bra{V} \nnum
\op S_{2} & = & \op n_{D} - \op n_{A} = \ket{H}\bra{V} +
\ket{V}\bra{H} \nnum
\op S_{3} & = & \op n_{R} - \op n_{L} = i(\ket{V}\bra{H} -
\ket{H}\bra{V})
\label{eq:STOKES}
\eea
where $\op n_{J}$ is the number operator for the $J$th polarization
mode. The eigenstates of $\op S_{1}$ are $\ket{H}$ and $\ket{V}$ with
eigenvalues $+1$ and $-1$ respectively. Similarly the
eigenstates of $\op S_{2}$ are $\ket{D}$ and $\ket{A}$ and of
$\op S_{3}$ are $\ket{R}$ and $\ket{L}$. The expectation values of
the Stokes operators are related to measurement by
\bea
\exptn{\op S_{1}} & = & {{2 R_{H}}\over{R_{H} + R_{V}}} - 1 \nnum
\exptn{\op S_{2}} & = & {{2 R_{D}}\over{R_{H} + R_{V}}} - 1 \nnum
\exptn{\op S_{3}} & = & {{2 R_{R}}\over{R_{H} + R_{V}}} - 1
\label{eq:STOKES2}
\eea
where $R_{H}$ and $R_{V}$ are the count rates recorded at the $H$
and $V$ output ports respectively of a horizontal/vertical
polarizing beamsplitter and similarly for the diagonal/anti-diagonal
and right/left bases.

On the other hand we can also express the expectation values in terms
of the density operator as
\bea
\exptn{\op S_{1}} & = & Tr[\op{\rho}\op S_{1}] =
\rho_{h,h} - \rho_{v,v} \nnum
\exptn{\op S_{2}} & = & Tr[\op{\rho}\op S_{1}] =
\rho_{h,v} + \rho_{v,h} \nnum
\exptn{\op S_{3}} & = & Tr[\op{\rho}\op S_{1}] =
i(\rho_{h,v} - \rho_{v,h})
\label{eq:STOKES3}
\eea
where $\rho_{i,j} = \bra{I}\op \rho \ket{J}$ are the elements of the
density matrix $\rho$ representing the density operator in the $H/V$
basis. Equations (\ref{eq:STOKES3})
are obtained using the ket representation of the Stokes
operators given in equation (\ref{eq:STOKES}). Combining equations
(\ref{eq:STOKES2}) and (\ref{eq:STOKES3}) we can obtain
all the elements of the density matrix in terms of the Stokes
operators expectation values and hence in terms of count rates. 
%via
%%
%\bea
%\rho_{h,h} & = & {{1+\exptn{\opt S_{1}}}\over{2}} =
%{{R_{H}}\over{R_{H} + R_{V}}} \nnum
%\rho_{v,v} & = & {{1-\exptn{\opt S_{1}}}\over{2}} =
%{{R_{V}}\over{R_{H} + R_{V}}} \nnum
%\rho_{v,h} & = & {{\exptn{\opt S_{2}}+i\exptn{\opt S_{3}}}\over{2}} =
%{{R_{D}+i R_{R}}\over{R_{H} + R_{V}}} - {{1 + i}\over{2}} \nnum
%\rho_{h,v} & = & {{\exptn{\opt S_{2}} - i \exptn{\opt S_{3}}}\over{2}} =
%{{R_{D} - i R_{R}}\over{R_{H} + R_{V}}} - {{1 - i}\over{2}}
%\eea
%%
%where we have used the normalization $\rho_{h,h}+\rho_{v,v}=1$. 
The
density matrix contains all the information about the polarization
state of the photons and properties such as the purity of the state
are readily extracted. A
question often asked is how similar the experimentally produced
state, $\op \rho$, is to
some pure target state $\ket{\phi}$. A common measure of this is the
fidelity, $F$, given by 
\beq
F = \bra \phi \op \rho \ket \phi
\label{eq:FIDELITY}
\eeq
which is easily found in terms of the matrix elements of $\rho$.
This technique can be extended to two, or more, photons by considering
the expectation values of products of Stokes operators of each
photon. For example
\beq
\exptn{\op S_{1,a} \op S_{1,b}} = \rho_{hh,hh} - \rho_{hv,hv} -
\rho_{vh,vh} + \rho_{vv,vv}
\eeq
where $a,b$ label the two photons and $\rho_{ij,kl} = \bra i \bra j
\op \rho \ket k \ket l$. By considering the expectation values of
all the different combinations of Stokes operator products the two
photon density matrix can be characterised. Whilst 4 measurements are
needed to completely characterise a single photon, 16 measurements are
needed in general for two photons. Of course if the two photons are
known to be in a separable state then 4 measurements on each
individual photon will suffice to characterise the state. The greater
number of measurements needed to characterise entangled states points
to their increased complexity. The exponential increase in
measurements required as a function of photon number
continues with three photons requiring 64
measurements and so on. 
%An example of the density matrix of a two qubit entangled state is shown in  Fig.\ref{fig4}

The above techniques were developed and applied by White
and James et al \cite{WHI99}, \cite{JAM01}. One problem that arises is
that, due to experimental errors, the density matrix produced from the
data may be unphysical. To deal with this, maximum likelihood
techniques were applied such
that the ``closest'' physical density matrix to the data can be
identified \cite{JAM01}.

%A large range of single and two photon states have now been analysed
%in this way with very
%high fidelity ($\ge 99\%$). In section (\ref{sec:QC}) we will see how
%these techniques can be applied to the characterization of quantum
%gates.
% A couple of example density matrices are shown in
% figure (\ref{fig:kwi}).
 %Figure. Figure ?
%
%\begin{figure}[htb]
%\begin{center}
%\includegraphics*[width=10cm]{progfig4}
%\caption{Density matrix in the horizontal/vertical basis of the two photon entangled state $\ket{HH}-\ket{VV}$}
%\label{fig4}
%\end{center}
%\end{figure}
 
\begin{figure}[htb]
\begin{center}
\includegraphics*[width=10cm]{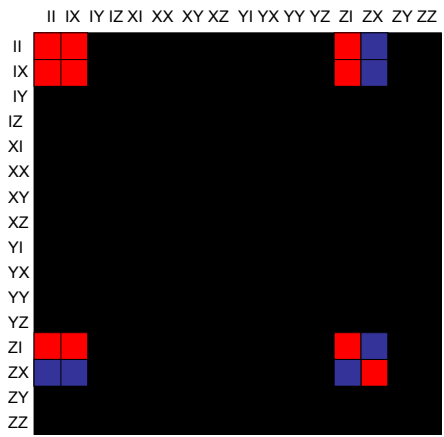}
\caption{Process matrix in the Stokes (equivalently Pauli) basis of the two qubit entangling gate the CNOT. Red = +0.25; yellow = -.25; black = 0.}
\label{fig5}
\end{center}
\end{figure}

 An unknown process can also be characterized by tomographic techniques \cite{NIE00}. An arbitrary single qubit process takes an input state $\op \rho$ to an output state $\op \rho'$. This can be written quite generally in terms of the Stokes operators as
 \begin{equation}
 \op \rho' = \sum_{i,j=0,3} w_{ij} \op S_i \op \rho \op S_j
\end{equation}
where we have introduced the Stokes identity operator: $\op S_0 = \ket{H} \bra{H} +  \ket{V} \bra{V}$. The coefficients, $w_{ij}$, form a matrix that completely describes the process. The process matrix can be determined from the expectation values of the Stokes operators evaluated for a complete set of input states. As a result 16 mesurement settings are required for single qubit process tomography. these techniques can be generalized to multi-qubit processes with a corresponding exponential increase in the number of measurements required. The process matrix corresponding to a CNOT gate is shown in Fig.\ref{fig5}. An experimental demonstration of process tomography on a two qubit circuit has been implemented in optics by O'Brien and Pryde et al \cite{JOB04}. 

\section{Quantum Key Distribution}
\label{SEC:QKD}

Perhaps the most straightforward application of quantum information technology is in the
field of secure communications. It is referred to variously as
quantum key distribution (QKD), quantum cryptography or sometimes quantum
key expansion and was initially proposed by Bennett and Brassard
\cite{BEN84}. \index{cryptography}\index{QKD}
The idea is to set up a communication channel which is
secure in the sense that any attempt to eavesdrop on the
communication can be detected after the fact. The channel is used to
send a random number encryption key between two parties, usually
referred to as Alice (the sender) and Bob (the receiver). The parties
then check if an eavesdropper, called Eve, intercepted any information
about the key. If no Eve was present they can proceed to use the
random number key to encrypt secret messages. If they find an Eve is
present they scrap that key and try again.

Actually in any practical situation there will always be some errors
in the transmission due to imperfections in the system. Thus what
Alice and Bob do in practice is to set limits on the amount of information that Eve
can have obtained based on the error rate they observe. Provided this
error rate is sufficiently small, post
processing of the data using techniques called error reconciliation
and privacy amplification \cite{BEN95} can be used to produce a
shorter secret key. Eve's information about this shorter key can be made
vanishingly small. Another important caveat is that Alice and Bob must
initially share some secret information which they can use to identify
each other. Otherwise Eve can fool them by pretending to be Bob to
Alice and vice versa. Given these conditions QKD is provably secure
\cite{GOT03}. No comparable result exists for classical communications.

\subsection{QKD using Single Photons}
\label{sec:QKDSP}

QKD's ability to detect eavesdroppers is based on the fact that any
process which acquires information about an observable of a quantum mechanical
system inevitably disturbs the values of other non-commuting observables.
To illustrate this consider the situation in which Alice is trying to communicate zero's and one's to Bob using polarized photons. First suppose Alice sends out a ``zero'' encoded as a
horizontally polarized photon, $|H \rangle$. Eve measures in the
horizontal/vertical basis and obtains
the result ``zero'' and so sends a horizontally polarized photon on
to Bob who will definitely get a zero if he measures in the
horizontal/vertical basis. However, now suppose Alice and Bob switch
to encoding in the diagonal/anti-diagonal basis without Eve knowing.
Alice sends a zero as a diagonally polarized photon,
$|D \rangle=(1/\sqrt{2})(|H \rangle + |V \rangle)$. Eve still measures in
the horizontal/vertical basis and so has a 50/50 possibility of getting
either zero or one as the result, regardless of what Alice sent.
Further more, what Eve sends on to Bob is basically the mixed state
$\rho = 1/2(|H \rangle \langle H|+|V \rangle \langle V|)$. So when Bob
measures in the diagonal/anti-diagonal basis he also gets a random
result. Thus, by measuring in the wrong basis, not only does Eve
potentially get the wrong result, but she also completely erases the
qubit value which is sent on to Bob who then may also get the wrong
result. 

The trick then is to arrange a situation in which Eve does not know
in which basis the information on any particular photonic qubit has
been encoded because then she is bound to make mistakes which Bob
will be able to detect. A typical protocol would go as follows:
\begin{enumerate}
\item
Alice sends a random number sequence to Bob, encoded on the
polarization of single photons. She randomly swaps between encoding
on the horizontal/vertical basis and encoding on the
diagonal/anti-diagonal basis.

\item
Bob measures the polarization of the
incoming photons and records the results, but he also swaps randomly
between measuring in the horizontal/vertical basis and measuring in
the diagonal/anti-diagonal basis.

\item
After the transmission is complete Alice and Bob communicate on a
public channel. First Bob announces which basis he measured for each
transmission event. Alice tells him whether or not this corresponded
to the basis in which she prepared the photon. They discard all
transmission events for which their bases did not correspond.

\item
Bob then reveals the bit values he measured for a randomly selected
subset of the remaining data. Alice compares the values revealed by
Bob with those she sent. Inevitably there will be some errors in the transmission. If this error rate is below a certain threshold then reconcilliation and privacy amplification can be employed to distill a secret key. If there are too many errors the data is
discarded and they try again.

\end{enumerate}

The first experimental demonstration of QKD was carried out by Bennett
and co-workers in 1992 over a distance of centimetres \cite{BEN92a}.
Demonstrations over distances of tens of kilometers were first carried out by Hughes et al in free space \cite{HUG02} and by Gisin et al in
fibre \cite{STU02}. Typically highly attenuated lasers
are used as the qubit source. Switching between the four input states
may be achieved through electro-optic control or via the passive
combination of four separate laser sources. In all cases it is
crucial that the spatial and temporal modes of the four input states
are identical so that no additional information is leaked to Eve. The
receiver station can be a passive arrangement. A 50/50 beamsplitter
is used to randomly send the incoming photons either to a
horizontal/vertical analyser or a diagonal/anti-diagonal analyser.
% A
%schematics of the QKD transmitter and receiver used in the free space
%demonstration by R.Hughes et al.  at Los Alamos Nation Laboratory
%is shown in figure \ref{fig:hughes}.

%%
%%
%%Figure. Figure 2 from HUG02
%\bef
%\vspace{50mm}
%%\leavevmode \epsfbox{Ou92Fig1.eps}
%\caption{\it Laser and detector arrangement used in free space QKD
%experiments of R.Hughes et al.\cite{HUG02}}
%\label{fig:hughes}
%\eef

To increase the signal to noise of the detection system the
detectors are gated, only opening for the one nanosecond or so window
in which the single photon pulse is expected. Synchronization may be
arranged via bright timing pulses preceding the single photon pulses
or via more standard public communication links. Sophisticated
reconciliation and privacy amplification algorithms then need to be
implemented over the public channel. %In the Hughes et al. experiment
%secure key rates of hundreds per
%second were achieved over a ten  kilometer range in broad daylight, a
%significant achievement. Somewhat higher rates were achieved at night.

The main motivation for free space systems is to transmit secret keys
to satellites securely. For terrestrial systems transmission through
fibre optic networks is more desirable. Although this has the
advantage of less stray light, it has the problem that optic fibre is
birefringent and hence polarization encoded qubits can become scrambled.
One solution is to go to the temporal mode qubit encoding. For example
one could use the non-commuting encodings
\bea
\ket{\bf 0} & \equiv & \ket{T1} + \ket{T2} \nnum
\ket{\bf 1} & \equiv & \ket{T1} - \ket{T2}
\eea
and
\bea
\ket{\bf +} & \equiv & \ket{T1} + i \ket{T2} \nnum
\ket{\bf -} & \equiv & \ket{T1} - i \ket{T2}
\eea
where $\ket{Ti}$ represents a single photon occupying a temporal wave
packet centred at time $Ti$. Alice could produce the state
$\ket{T1} + \ket{T2}$ by allowing a single photon pulse to pass through a Mach
Zehnder interferometer with unequal arm lengths,
in particular where the arm length
difference is $T1-T2$. The other states are created in a similar way
but where an additional phase of $\pi$ (for the state
$\ket{T1} - \ket{T2}$) or $\pi/2$ or $3 \pi/2$ (for the
other basis states) is added to one arm of the interferometer.
Unfortunately, a readout by Bob would require him to have an
interferometer which is phase-locked to Alice's , something that is 
difficult to arrange.
An elegant solution to this
problem is for Bob to first send a bright pulse to Alice \cite{STU02}.
This pulse
acts as a phase reference, thus avoiding the locking issue. %The 
%details are shown in
%Fig.(\ref{fig:zbi}). Using this arrangement they were able to transmit
%secure keys over distances of nearly 70 [km] at rates of about
%50 [per s].

%
%Figure. Figure 2 from STU02
%\bef
%\vspace{50mm}
%%\leavevmode \epsfbox{Ou92Fig1.eps}
%\caption{\it Schematics of the experiment by D.Stucki et al.\cite{STU02}.
%A strong laser pulse, emitted by Bob is first split on a beamsplitter
%(BS). The two pulses are recombined on a polarising beamsplitter (PBS)
%after passing through a short arm and a long arm including a phase
%modulator ($PM_{B}$) and a 50[ns] delay line (DL). The polarisation
%of the short arm is rotated 90 degrees such that both pulses exit by
%the same port of the PBS. The pulses travel to Alice where they are
%reflected on a Faraday mirror, are attenuated and are sent back to
%Bob orthogonally polarized. As a result the pulses return along their
%paths in reversed order and interfere at BS. The entire system forms
%an autocompensated interferometer. The ``0'' , ``1'' basis is encoded by
%Alice applying either $0$ or $\pi$ phase shift with $PM_{A}$ and read
%out by Bob with no phase shift applied to the return pulse.
%The ``+'' ``-'' basis is encoded by
%Alice applying either $\pi/2$ or $3\pi/2$ phase shift with $PM_{A}$ and read
%out by Bob with $\pi/2$ phase shift applied via $PM_{B}$ to the return pulse.}
%\label{fig:zbi}
%\eef
%

A limit on the secure key rates occurs with the use of attenuated
laser sources. The initial intensity cannot be too great otherwise the
probability of two-photon events will be too high. Eve can use two
photon events to extract information about the key without penalty.
One solution to this problem is to use a true single photon source (see section \ref{sec:TSPS}).
Beveratos et al. \cite{BEV02} were the first to demonstrate such a
scheme. They used the fluorescence from a single NV colour centre
inside a diamond nano-crystal at room temperature as their single
photon source. 
%The nano-crystals were attached to a dielectric mirror
%and their emission was collected by a high numerical aperture
%microscope objective. The effective value of $\tau$ (see equation
%(\ref{TAU})) for this source was $0.07$, thus the number of two photon
%events is reduced by a factor of $14$ over an attenuated laser source
%of the same intensity. A test system employing this source was able to
%achieve secure bit rate of $7,700$ [per s] over 50 [m] in
%free space.

The QKD protocol we have discussed here is called BB84.
Many other protocols have been
proposed and demonstrated and new protocols and
demonstrations appear regularly \cite{GIS02}. Initial steps to
commercialization have already been taken.

\subsection{QKD using Continuous Variables}

An alternative approach to QKD is to use non-commuting continuous variables such as the in-phase and out-of-phase quadratures. We saw in section \ref{CIL} that information can be encoded on the quadrature amplitudes and read out using homodyne detection. We now examine the use of these techniques for QKD.

Recall the basic mechanism used in QKD schemes is the fact
that the act of measurement (by Eve) inevitably
disturbs the system. This measurement back-action of course also exists for
continuous quantum mechanical variables. In particular let us
consider the situation in which Alice sends a series of weak coherent states
to Bob whose amplitudes are picked from a two-dimensional Gaussian
distribution centred on zero. Bob chooses to measure either the
in-phase or out-of-phase projections of the states
onto a shared local oscillator
using homodyne detection.
Bob will effectively
see a Gaussian distribution of real amplitude coherent states when he
looks in-phase and a Gaussian distribution of imaginary amplitude
coherent states when he looks out-of-phase. Alice can encode
two different random number sequences on the two. Because the two
quadrature measurements do not commute Eve now
has a similar problem as in the discrete case: any attempt to extract
information about one quadrature from the beam will inevitably erase
information carried on the other quadrature. If Alice and Bob compare
some of the data at the end of the protocol they will thus notice
increased error rates as a result of any intervention. This
protocol was developed by Grosshans and Grangier \cite{GRO02} based on
earlier work \cite{RAL00,CER01} and has been developed considerably since \cite{SIL02,WEE04,GRO05}. Security proofs for coherent state protocols are based on various reasonable assumptions however, a proof of absolute security on par with those made in the discrete case has only been made for a
somewhat different continuous
variable protocol based on squeezed states \cite{GOT02}.

Coherent state QKD can be implemented either by sending very weak
coherent pulses of light or by sending bright, quantum limited light
with in-phase or out-of-phase amplitude modulation playing the role of
the coherent states.
The first in principle experimental
demonstration of this technique was performed by Grosshans et al.
\cite{GRO03} using the former technique. Recently the latter technique has been employed \cite{LOR04,LAN05}. In the experiment of Lance et al \cite{LAN05} end to end key exchange in the presence of 90\% loss was achieved with a secret key rate of 1Kbit/s with a 17MHz bandwidth.  Since this scheme is truly broadband, it can potentially deliver orders of magnitude higher key rates by extending the encoding bandwidth with higher-end telecommunication technology. 

%
%\subsection{No Cloning}
%\label{sec: NO CLONING}

%An alternative way of understanding the basic mechanism of QKD is the
%fact that quantum information cannot be copied or {\it cloned}.
%That is, if given an unknown quantum state, it is not
%possible to produce an identical copy or clone of it. This was first
%pointed out by W.K.Wootters, W.H.Zurek \cite{WOO82}. The best that can be
%achieved is to turn the original input state into a pair of output
%states which are approximate copies of the original. For qubits the
%best achievable fidelity of the clones with the original is $5/6$.
%For coherent states the best achievable fidelity is $2/3$.
%\index{clone} \index{no-cloning limit}

%\section{Entanglement}
%\label{sec:ENTANGLEMENT}

%
%\section{Quantum State Sharing}

%
\section{Quantum Teleportation}
\label {sec:TELEPORTATION} 

We have seen that quantum communication can be more secure than classical communication. When entangled states are allowed a number of new enhanced communication and processing tasks become possible.
This is quite remarkable given that
entanglement is undirected and carries no information itself. For example, in the presence of entanglement, the classical capacity of a quantum channel (ie the ability of a quantum system to carry classical information) is increased. This is called  {\it quantum dense
coding} and has been described both in the discrete  \cite{BEN92b} and continuous domains \cite{BRA00,RH02}. In principle experimental demonstrations have also been made in both domains \cite{MAT96,PAN00}. Another example is quantum state sharing. Here an unknown quantum state can be distributed between $n$ parties in such a way that if $m$ of the parties collaborate (where $m<n$), then the state can be retrieved, but less than $m$ parties can not retrieve the state. Again both discrete \cite{CLE99} and continuous \cite{TYC03} protocols are known and an experimental demonstration has been made in the continuous case \cite{LAN04}.
%example, suppose Alice and Bob share a pair of qubits in the entangled
%state $\ket{0}_{a}\ket{0}_{b}+\ket{1}_{a}\ket{1}_{b}$. Alice chooses
%to do one of four operations on her qubit:
%%
%\begin{enumerate}
%\item
%Leave the qubit unchanged.

%\item
%Perform a bit-flip or $X$ operation, which takes $\ket{0} \to \ket{1}$ and
%$\ket{1} \to \ket{0}$.

%\item
%Perform a phase-flip or $Z$ operation, which leaves $\ket{0}$ unchanged but
%takes $\ket{1} \to -\ket{1}$.

%\item
%Perform a bit-flip and a phase-flip.

%\end{enumerate}
%%
%She then sends her qubit to Bob who then has a pair of qubits in one
%of the four states:
%%
%\begin{enumerate}
%\item
%$\ket{\phi+} = \ket{0}_{a}\ket{0}_{b}+\ket{1}_{a}\ket{1}_{b}$,

%\item
%$\ket{\psi+} = \ket{1}_{a}\ket{0}_{b}+\ket{0}_{a}\ket{1}_{b}$,

%\item
%$\ket{\phi-} = \ket{0}_{a}\ket{0}_{b}-\ket{1}_{a}\ket{1}_{b}$,

%\item
%$\ket{\psi-} = \ket{1}_{a}\ket{0}_{b}-\ket{0}_{a}\ket{1}_{b}$,

%\end{enumerate}
%%
%respectively. These states are known as the Bell-states. The important
%point is that they are mutually orthogonal and so Bob can now, in
%principle, perform a measurement which unambiguously picks which of the
%four states he has (this is known as a Bell measurement). The outcome
%is that Alice has communicated two bits of information to Bob whilst
%only sending a single qubit! This effect is called {\it quantum dense
%coding }, or super dense coding, \cite{BEN92b} and has no classical
%analogue.
%\index{dense coding}\index{X operation}\index{Z operation}
%\index{state!Bell}\index{Bell!state}\index{Bell!measurement}
Perhaps the most surprising (and most useful) of such tasks is quantum teleportation \cite{BEN93}.
Here the presence of entanglement enables Alice to send Bob an unknown
qubit by simply sending a classical message. To understand the novelty of this consider first what
Alice can do in the absence of entanglement. Her best strategy is
to measure the qubit in some basis and send a message that tells
Bob to make a qubit
corresponding to the result she gets. Sometimes she will be lucky and
will measure in a good basis, then Bob will make a close approximation
to the qubit.
Other times she will measure in a bad basis and the result she
gets will be completely random and Bob will make a poor approximation
to the qubit. It can be shown that on average the fidelity of Bob's
qubit with Alice's original is $2/3$.

If they share entanglement they can perform teleportation. This
works in the following way: Alice and Bob share an
entangled pair of qubits, say
$\ket{0}_{a}\ket{0}_{b}+\ket{1}_{a}\ket{1}_{b}$ where the subscripts indicate the party that has the qubit. Alice also has a
qubit in the arbitrary state $\mu \ket{0}_{a'} +\nu \ket{1}_{a'}$
which she wishes to send to Bob. Alice does not know the state of her
qubit. If we write down the state of the three qubits and then
rearrange it we notice a remarkable feature: Bob's qubit can be
represented as an equal superposition of four states, each differing
from Alice's unknown qubit by at most a bit-flip and a phase-flip.
\bea
(\mu \ket{{\bf 0}}_{a'} &+&\nu \ket{{\bf 1}}_{a'})
(\ket{{\bf 0}}_{a}\ket{{\bf 0}}_{b}+\ket{{\bf 1}}_{a}\ket{{\bf 1}}_{b})\nnum
 &=&   (\ket{{\bf 0}}_{a'}\ket{{\bf 0}}_{a}+\ket{{\bf 1}}_{a'}\ket{{\bf 1}}_{a})
\;\; (\mu \ket{{\bf 0}}_{b} +\nu \ket{{\bf 1}}_{b}) \nnum
& + & \;(\ket{{\bf 0}}_{a'}\ket{{\bf 0}}_{a}-\ket{{\bf 1}}_{a'}\ket{{\bf 1}}_{a})
\;\; (\mu \ket{{\bf 0}}_{b} -\nu \ket{{\bf 1}}_{b}) \nnum
& + & \;(\ket{{\bf 0}}_{a'}\ket{{\bf 1}}_{a}+\ket{{\bf 1}}_{a'}\ket{{\bf 0}}_{a})
\;\; (\mu \ket{{\bf 1}}_{b} +\nu \ket{{\bf 0}}_{b}) \nnum
& + & \;(\ket{{\bf 0}}_{a'}\ket{{\bf 1}}_{a}-\ket{{\bf 1}}_{a'}\ket{{\bf 0}}_{a})
\;\; (\mu \ket{{\bf 1}}_{b} -\nu \ket{{\bf 0}}_{b}) \nnum
\;
\eea
What is more, each of Bob's outcomes corresponds to distinct 2 mode states on Alice's side. Thus Alice can tell which of these four states Bob actually has
by making measurements in the so-called {\it Bell basis} of her two qubits, explicitly:
\begin{enumerate}
\item
$\ket{\phi^+} = \ket{{\bf 0}}_{a}\ket{{\bf 0}}_{a'}+\ket{{\bf 1}}_{a}\ket{{\bf 1}}_{a'}$,

\item
$\ket{\psi^+} = \ket{{\bf 1}}_{a}\ket{{\bf 0}}_{a'}+\ket{{\bf 0}}_{a}\ket{{\bf 1}}_{a'}$,

\item
$\ket{\phi^-} = \ket{{\bf 0}}_{a}\ket{{\bf 0}}_{a'}-\ket{{\bf 1}}_{a}\ket{{\bf 1}}_{a'}$,

\item
$\ket{\psi^-} = \ket{{\bf 1}}_{a}\ket{{\bf 0}}_{a'}-\ket{{\bf 0}}_{a}\ket{{\bf 1}}_{a'}$,

\end{enumerate}
If the result of Alice's Bell measurement is $\ket{\phi^+}$ she tells
Bob not to do anything, if it is $\ket{\phi^-}$ she tells him to do a
phase-flip, if it is $\ket{\psi^+}$ a bit-flip is required and finally
if she measures $\ket{\psi^-}$ she tells him to perform both a bit and
a phase-flip. In the end Bob has turned his qubit into an exact copy of
Alice's original but all Alice has sent is a two bit classical message.

\subsection{Teleportation of single photon qubits}
\label{sec:TELEQB}

The key ingredients for a demonstration of teleportation are the ability to produce entangled Bell States and to measure in the Bell basis. Both of these things can be achieved non-deterministically in optics. Down conversion (see section \ref{sec:OQ}) can be run in such a way as to produce polarization entangled photon pairs. As shown by Kwiat el al, this can be achieved either in a type I \cite{WHI99} or type II \cite{KWI95} scenario. In both cases the basic idea is to overlap output modes in such a way that the photon pairs can either be comprised of two horizontal photons or two vertical photons but are otherwise completely indistinguishable (for type II the polarisations are anti-correlated). This leads to an output state that can be written
\begin{equation}
\ket{0} \ket{0} + \chi (\ket{H}_{a}\ket{H}_{a'}+ e^{i \theta} \ket{V}_{a}\ket{V}_{a'}) + ....
\end{equation}
where we assume $\chi<<1$ and so ignore higher order terms. The phase $\theta$ can be tuned experimentally. By tuning $\theta=0$ and working in coincidence, the maximally entangled Bell state, $\ket{\phi^+} $, can be post-selected. 

\begin{figure}[htb]
\begin{center}
\includegraphics*[width=9cm]{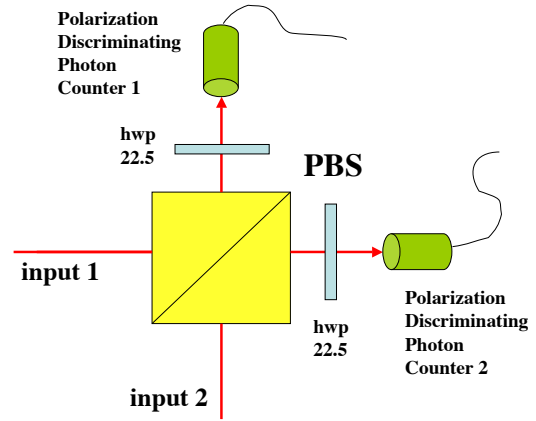}
\caption{Schematic of a partial Bell state analyser. The polarization discriminating detectors would be constructed from additional polarizing beamsplitters (PBS) and photon counters in practice. If we find a single photon at each output and the polarization of the photons at each output are the same then the Bell state $\ket{\phi^+} $ has been identified. If the polarization of the photons at each output are different then $\ket{\phi^-} $ has been identified. If both photons are found at the top detector then the separable state $\ket{HV}$ has been identified. If both photons are at the bottom port then the separable state $\ket{VH}$
 has been identified. hwp is a half-wave plate at the angle indicated.}
\label{fig6}
\end{center}
\end{figure}

Partial Bell state measurements in the polarization basis can be achieved with the beamsplitter and photon counter arrangement shown in Fig.\ref{fig6} \cite{WEI94,BRA95}. The measurement projects onto the basis states $\ket{\phi^+} $, $\ket{\phi^-} $, $\ket{HV}$, $\ket{VH}$. We see that in half the cases we project onto Bell states and so can achieve teleportation. The other half of the time we make a separable measurement of the individual values of the qubits, and so teleportation fails.

The first demonstration of teleportation using these resources was by Bouwmeester et al.
at the University of Innsbruck \cite{BOU97}.
Their experiment used pulsed UV pumping 
of a non-linear crystal in a
type II arrangement to produce pairs of polarization entangled
photons at 788[nm] (2 and 3). The pump pulse was then retro-reflected through
the crystal such that a second pair of counter-propagating photons (1
and 4) might be produced. Teleportation could then proceed by giving
Alice entangled photon 2 and photon 1 as the teleportee (after it had
been prepared in some arbitrary state) and giving Bob entangled photon 3. The fourth photon could be used as a trigger.

A simpler form of the partial Bell measurement was implemented by passing
photons 1 and 2 through a 50/50 beamsplitter and then photon counting at the
outputs. The action of a beamsplitter on the Bell states is
to make the photons bunch (i.e. both exit through the same port). For all the Bell states that is
except $\ket{\psi-}$, for which case the photons always exit by
different ports. Thus if Alice records a coincidence count at the
output of the beamsplitter then she has unambiguously identified the
$\ket{\psi-}$ Bell state and teleportation has succeeded. The
experiment is arranged such that the $\ket{\psi-}$ state is the ``do
nothing'' result. If she does not
obtain a coincidence the protocol has failed.
%
%Results from the experiment are shown in Fig.(\ref{fig:boures}).
%Teleportation should have succeeded when: photon 4 is detected at p
%(indicating that photon 1 is on its way); a coincidence is detected at
%f1 and f2 (indicating the correct Bell state has been detected) and; a
%photon is detected at either d1 or d2 (indicating the teleported
%photon has been successfully detected). When these 4-fold coincidences
%occur we should expect the polarisation state of Bob's photon to match
%that of the photon prepared for Alice. This is what is shown in
%Fig.(\ref{fig:boures}) for input photons in the non-orthogonal
%states $\ket{A}$ and $\ket{H}$ (-45 and 0 degrees respectively). At
%zero time delay between when photons 1 and 2 meet at their
%beamsplitter the original states are reproduced with a visibility
%of $70\% \pm 3\%$. When a time delay is introduced, photons 1 and 2
%become distinguishable at the beamsplitter, the Bell measurement
%fails and so too does the teleportation.

%

%%
%%Figure. Figure 5 from BOU97
%\bef
%\vspace{50mm}
%%\leavevmode \epsfbox{Ou92Fig1.eps}
%\caption{\it Experimental results from the Bouwmeester et al experiment
%\cite{BOU97}.}
%\label{fig:boures}
%\eef
%

This experiment was very technically challenging. The probability of
four photon events was very low. To prevent any
temporal distinguishability of photons 1 and 2, a frequency filtering 
producing a 4[nm]
bandwidth was applied, further reducing the counts. Finally
the protocol itself only succeeded one quarter of the time. This
resulted in roughly one successful event per minute. The fidelity with which the original states were reproduced was about $70\%$. 

A sublety of the
original experiment was that there was a significant probability for the
down-converter to produce two photons each in modes 1 and 4. Then, even
under perfect conditions of zero loss, a three-fold coincidence on Alice's
side of the experiment does not guarantee a photon is sent to Bob. In
a later manifestation of the experiment the possibility of such errors
was made negligible and fidelities of $>80 \%$ were observed, well in
excess of the $2/3$ limit \cite{PAN03}.

Teleportation can also be performed on single rail qubits. Here the Bell state can be produced by simply splitting a single photon on a 50:50 beamsplitter to give $\ket{0}\ket{1} + \ket{1}\ket{0}$. A Bell measurement is achieved also with a 50:50 beamsplitter and can successfully identify the two Bell states  $\ket{0}\ket{1} \pm \ket{1}\ket{0}$. Again the other possibilities (two photons at one output) result in the measurement of the logical value of the qubit. A demonstration of single rail teleportation was carried out by Lombardi et al \cite{LOM02}.

Teleportation of coherent state qubits is also possible and has the unique property that deterministic Bell state analysis can be carried out with just a beamsplitter \cite{ENK02,JEO02} (provided the coherent states are sufficiently separated to be considered orthogonal, see section \ref{subsec:SRE}). No experimental demonstration of this type of teleportation has yet been carried out.

As well as a method for quantum communication, all these types of teleportation can also be applied to quantum computation as will be highlighted in section \ref{sec:QC}.

\subsection{Continuous Variable Teleportation}
\label{sec:CVTELE}

So far we have considered teleportation of qubits, as carried by the
polarization degree of freedom of
single photons. This technique will only work for single photon
states. What if we wish to teleport a general field state with contributions
from vacuum and higher photon number terms? The answer is to
implement a teleportation protocol based on the measurement of the
quadrature amplitudes of the field. Because the quadrature amplitudes
are continuous, rather than
discrete, variables, this is known as continuous variable (CV)
teleportation. It was developed by Braunstein and Kimble
\cite{BRA98} based on
earlier work by Vaidman \cite{VAI94}.

\begin{figure}[htb]
\begin{center}
\includegraphics*[width=9cm]{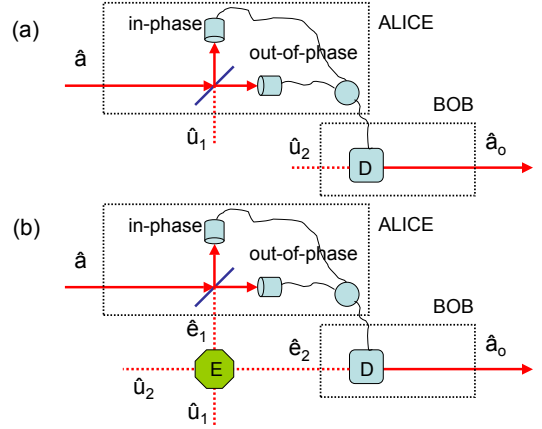}
\caption{Schematic of continuous variable teleportation. In (a) is depicted the best strategy in the absence of entanglement. In (b) entanglement (E) is included in the protocol. Homodyne detection measures the in-phase and out-of-phase quadratures at Alice's station. The results of the measurements are fed-forward to displace (D) Bob's field.}
\label{fig7}
\end{center}
\end{figure}

Consider the situation depicted in Fig.\ref{fig7}(a). Alice wishes to teleport to Bob an
unknown coherent state, $\ket{\alpha}$, drawn from
a broad Gaussian distribution. In the absence of entanglement Alice's
best approach is to divide the field into two equal parts at a beamsplitter
and then measure the in-phase quadrature of one half and the out-of-phase
quadrature of the other. The in-phase measurement gives an estimate
of the real part of $\alpha$, whilst the out-of-phase measurement gives an
estimate of the imaginary part of $\alpha$, however both estimates are
imperfect due to noise from the vacuum field which inevitably enters
through the open port of the beamsplitter. Alice sends these estimates
to Bob who uses them to produce a coherent state by displacing his local
vacuum state by the relevant quantities.

This situation is most easily described in the Heisenberg picture. Let
the initial field mode be represented by the annihilation operator
$\op a$ and the vacuum entering at the 50:50 beamsplitter by $\op
u_{1}$. The measurement results obtained by Alice are then
represented by the quadrature operators
\bea
\op{X}^+_{a} & = & {{1}\over{\sqrt{2}}}(\op{X}^+_{a} - \op{X}^+_{u1}) \nnum
\op{X}^-_{a} & = & {{1}\over{\sqrt{2}}}(\op{X}^-_{a} + \op{X}^-_{u1})
\eea
where $+$ ($-$) signifies the in-phase (out-of-phase) quadrature. These are sent to Bob who uses them to displace his vacuum field, $\op
{u}_{2}$ giving the output field
\beq
\op a_o = \op{u}_{2} + g {{1}\over{2}} (\op{X}^+_{a} +
i \op{X}^-_{a}) - g {{1}\over{2}} (\op{X}^+_{u1} -
i \op{X}^-_{u1})
\eeq
where $g$ is a gain factor for the displacement. Choosing
$g = 1$, unity gain, Bob's output field is
\beq
\op a_o = \op a + \op{u}_{2} - \op{u^{\dagger}}_{1}
\label{eq:ctele}
\eeq
Notice that two vacuum fields have been added to the output, one
entering through Alice's measurement, the other through Bob's
reconstruction. Measurement of Bob's quadrature amplitudes will show the same average value as Alice's input: $\langle \op X^\pm_{ao} \rangle = \langle \op X^\pm_a \rangle$ as the vacuums have zero mean. On the other hand the quadrature variances of Bob's state will be larger than the initial state:
\bea
V_{ao} & = & \langle (\op X^\pm_{ao})^2 \rangle - \langle \op X^\pm_{ao} \rangle^2 \nnum
& = & \langle (\op X^\pm_a)^2 \rangle + \langle (\op X^\pm_{u1})^2 \rangle + \langle (\op X^\pm_{u2})^2 \rangle- \langle \op X^\pm_a \rangle^2 \nnum
& = & 3
\eea
As a result Bob's state is mixed (no longer minimum uncertainty) and is 3 times
noisier than the QNL level of the input coherent state.

Now suppose Alice and Bob share an entangled state. In particular we assume they share an EPR entangled state \cite{REI88,OU92} named after the famous paradox proposed by Einstein, Podolsky and Rosen \cite{EIN35}. This state, also commonly known as a two mode squeezed state, exhibits strong correlations between both the in-phase and out-of-phase quadratures of its component beams. It can be described by its Heisenberg evolution:
\bea
\op{u}_{1} & \to & \sqrt{G} \;  \op{u}_{1} +\sqrt{G-1} \;  \op{u}_{2}^{\dagger} \nnum
 \op{u}_{2} & \to & \sqrt{G} \;  \op{u}_{2} +\sqrt{G-1} \;  \op{u}_{1}^{\dagger}
\label{eq:EPR3}
\eea
where $ \op{u}_{i}$ are initially vacuum states and $G$ is the parametric gain (or squeezing). This interaction is produced by parametric
amplification \cite{WU87} either directly by a non-degenerate
system or alternatively via degenerate parametric amplification
(ie squeezing) followed by the out-of-phase mixing of the two modes on a 50:50 beamsplitter. A non-degenerate parametric amplifier is basically just a high efficiency down converter as can be seen from the Scr{\"o}dinger picture evolution equivalent to Eq.\ref{eq:EPR3}:
\bea
\ket{EPR} &  = & {{1}\over{\sqrt{G}}} \;\; ( \ket{0}_{1} \ket{0}_{2} +
{{\sqrt{G-1}}\over{\sqrt{G}}} \ket{1}_{1} \ket{1}_{2} \nnum
& + &
\left({{\sqrt{G-1}}\over{\sqrt{G}}}\right)^{2}
    \ket{2}_{1} \ket{2}_{2} + \ldots.. )
\eea
Alice again divides and measures her beam but this time instead of
allowing vacuum to enter the empty port of her beamsplitter she sends
in her half of the EPR pair. As a result her quadrature measurement
results are now given by
\bea
\op{X}^+_{a} & = & {{1}\over{\sqrt{2}}}(\op{X}^+_{a} - \sqrt{G} \;
\op{X}^+_{u1} - \sqrt{G-1} \; \op{X}^+_{u2}) \nnum
\op{X}^-_{a} & = & {{1}\over{\sqrt{2}}}(\op{X}^-_{a} + \sqrt{G} \;
\op{X}^-_{u1} - \sqrt{G-1} \; \op{X}^-_{u2})
\eea
where we have used equation (\ref{eq:EPR3}) to describe the
entanglement. These results are sent to Bob who now uses them to
displace his half of the EPR pair, obtaining (at unity gain)
\beq
\op a_o = \op a + (\sqrt{G} - \sqrt{G-1}) \op{u}_{2} +
(\sqrt{G} - \sqrt{G-1}) \op{u^{\dagger}}_{1}
\label{eq:tele}
\eeq
Now in the limit $G \to \infty$, $(\sqrt{G} - \sqrt{G-1}) \to 0$,
hence in this limit equation (\ref{eq:tele}) reduces to
\beq
\op a_o = \op a
\eeq
Evolution through the teleporter is the identity and so the output
state is identical to the input (this is obviously true not only for the
coherent input states we have been considering but for any input state).

The first demonstration of this type was made by Furusawa et al.  \cite{FUR98}.
EPR entanglement was produced by the mixing of two
out-of phase squeezed beams on a 50/50 beamsplitter. Both squeezed 
beams (at 860[nm])
were generated in a single ring-cavity parametric oscillator by simultaneously
pumping counter-propagating cavity modes. One of the EPR beams was sent to Alice who mixed it with her
signal beam and performed dual balanced homodyne measurements, actively locked
to be 90 degrees out of phase, such that conjugate quadrature
measurements were made. The photo-currents thus generated are sent to
Bob who uses them to impose phase and amplitude modulations on a
bright laser beam. By mixing this bright beam with his EPR beam
on a highly reflective beamsplitter Bob can efficiently impose on the EPR beam
a displacement proportional to the
modulations. All the beams in the experiment originate from a single
Ti:Sapph master laser, including the signal beam which has a known
modulation amplitude (effectively the coherent amplitude of the coherent state) imposed on it before being sent to
Alice. When, based on the
signal size observed on Bob's side, unity gain was achieved, the
quadrature noise floors of the teleported beam were measured by an
independent balanced homodyne detector, both with
and without entanglement.

%
%%
%%Figure. Figure 1 from FUR98
%\bef
%\vspace{50mm}
%%\leavevmode \epsfbox{Ou92Fig1.eps}
%\caption{\it Schematical layout of the Caltech teleportation experiment.
%After A.Furusawa et al.\cite{FUR98}.}
%\label{fig:furset}
%\eef
%

The quality of Bob's reconstruction can be evaluated
via the fidelity of it compared with the initial coherent state Alice
sent (see Eq.\ref{eq:FIDELITY}). Provided the output is Gaussian (which it is) this fidelity is
given by
\bea
F & = & {{2}\over{\sqrt{(V^+_{ao}+1)(V^-_{ao}+1)}}}\nnum
& \times & \exp \left
[-{{2}\over{\sqrt{(V^+_{ao}+1)(V^-_{ao}+1)}}}|\alpha|^{2}(1-g)^{2}
\right ]
\label{eq:F}
\eea
Recalling from our earlier discussion that without entanglement the
quadrature variances of the outputs are $V^+_{ao} = V^-_{ao} = 3$, then we
find from equation (\ref{eq:F}) that for large $\alpha$ the best fidelity
with no entanglement is achieved at unity gain and is $F = 0.5$. This
is confirmed by Furusawa et al who find a best fidelity without
entanglement of $F_{c} = 0.48 \pm 0.03$. On the other hand, with
entanglement, a fidelity of $F_{q} = 0.58 \pm 0.02$ is measured,
clearly exceeding the classical bound. 

%
%
%Figure. Figure 5 from FUR98
%\bef
%\vspace{50mm}
%%\leavevmode \epsfbox{Ou92Fig1.eps}
%\caption{\it Measured fidelities from the experiment
%by A.Furusawa et al.\cite{FUR98}.
%Results with and without entanglement are shown along with
%theoretical curves.}
%\label{fig:furres}
%\eef
%

A subsequent experiment by Bowen et al  \cite{BOW03}.
achieved higher
fidelities ($F_{q} = 0.64 \pm 0.02$) and stable operation over long periods. Their experiment used
two independent, monolithic, sub-threshold parametric oscillators to produce twin
squeezed beams at 1064 [nm] which were then mixed on a beamsplitter to
produce the required EPR entanglement. 
%In figure
%(\ref{fig:bowres1}) spectra of the amplitude and phase quadratures of
%the input and teleported signals are shown along with time traces
%showing the stability of the system over several minutes.

%%
%%
%%Figure. Figure 2 from BOW03
%\bef
%\vspace{50mm}
%%\leavevmode \epsfbox{Ou92Fig1.eps}
%\caption{\it Input and output spectra for amplitude and phase
%quadratures from the ANU teleportation.
%Time traces show noise on amplitude and phase quadratures of the
%output over 5 [min] interval. From W.Bowen et al.\cite{BOW03}}
%\label{fig:bowres1}
%\eef
%

The performance of the Bowen et al
teleporter was also characterised in terms of the signal to noise transfer ($T$) and the conditional variance ($V$) between the input and output fields: the teleportation T-V diagram \cite{RAL98}.
As we have seen,
in the absence of entanglement strict bounds are placed on both the
accuracy of measurement and reconstruction of an unknown state.  These
are represented by the vacuum modes that appear in equation (\ref{eq:ctele}).
These bounds can be
quantified in the following way.

Alice's measurement accuracy is limited by the generalized uncertainty
principle of Arthurs and Goodman, $V^+_{M} V^-_{M} \!  \ge \!  1$  \cite{ART88},
where $V+_{M},V^-_{M}$ are the quadrature measurement penalties, which
holds for
{\it any} simultaneous measurements of conjugate quadrature
amplitudes of an unknown quantum optical system. For Gaussian input states this relationship can be re-written in terms of quadrature signal transfer
coefficients, $T^+={S/N}^+_{\rm out}/{S/N}^+_{\rm in}$ and 
$T^-={S/N}^-_{\rm out}/{S/N}^-_{\rm in}$ as
\begin{equation}
T_{q}=T^++T^--T^+\;T^- \left (1-\frac{1}{V^+_{in} V^-_{in} }\right ) \le 1
\label{tu}
\end{equation}
where ${S/N}^+$ (${S/N}^-$) is the signal-to-noise ratio of the in-phase (out-of-phase) quadratures.  This expression reduces to
$T_{q}=T^++T^-$ for minimum uncertainty input states ($V^+_{in} V^-_{in} =1$).  Without entanglement it is not possible
to break the inequality given in  equation (\ref{tu}).

Bob's reconstruction must be carried out on a mode of the E/M field
the fluctuations of which must already obey the uncertainty
principle.
In the absence of entanglement these intrinsic fluctuations remain
present on any reconstructed field, thus the amplitude and phase
conditional variances, $V^+_{{\rm in|out}} \!  = \!  V^+_{{\rm
out}}\!  - \!  |\langle \delta \hat X^+_{{\rm in}}
\delta  \hat X^+_{{\rm out}}\rangle|^{2} \!  /V^+_{{\rm
in}}$ and $V^-_{{\rm in|out}} \!  = \!  V^-_{{\rm
out}}\!  - \!  |\langle \delta \hat X^-_{{\rm in}}
\delta  \hat X^-_{{\rm out}}\rangle|^{2} \!  /V^-_{{\rm
in}}$ respectively, which
measure the noise added during the teleportation process, will satisfy
$V_q=V^+_{{\rm in|out}} V^-_{{\rm in|out}} \!  \ge \!  1$.  For Gaussian input states this can
be written in terms of the signal transfer and quadrature variances of
the output state as
\begin{equation}
V_{q}=(1-T^+)(1-T^-) \; V^+_{\rm out} V^-_{{\rm
out}}\;\ge 1
\label{vtu}
\end{equation}
The criteria of equations (\ref{tu}) and (\ref{vtu})
can then used to represent 
quantum teleportation on a T-V graph. An important feature of the T-V criteria is that it can characterize teleportation at non-unity gains \cite{RAL99}.

The $T_{q}$ and $V_{q}$ bounds have independent physical significance.  If
Bob's state passes the $T_{q}$ bound (equation (\ref{tu})) then he can be sure,
regardless of how it was transmitted to him, that no other party can
possess a copy of the state which also passes this bound (ie carries
as much information about the original).
Surpassing the $V_{q}$ bound is a
necessary prerequisite for reconstruction of non-classical features
of the input state such as squeezing or negativity of the Wigner function.  Clearly it is desirable that
the $T_{q}$ and $V_{q}$ bounds are simultaneously exceeded, thus demonstrating fully quantum operation.  The cross-over point (1,1), corresponds to a fidelity of $2/3$. The significance of crossing this boundary has been investigated by a number authors \cite{GRO01,CAV04}.
Perfect reconstruction of the input state would result in $T_{q} \!  = \!  2$
and $V_{q} \!  = \!  0$. In the Bowen et al experiment a number of
results were obtained that passed the $T_{q}$ bound and one point (marginally) exceeded both
$T_{q}$ and $V_{q}$ simultaneously. 

More recently Takei et al \cite{TAK05} conclusively demonstrated passage into the fully quantum region by obtaining a fidelity of $0.7$ and realizing entanglement swapping at unity gain. The experiment was carried out with an array of 4 parametric oscillators operating at 860 [nm]. By combining pairs of beams, 2 strongly entangled EPR sources were created. One beam from the first EPR source was teleported by the second EPR source. By looking for correlations between the teleported beam and the other beam of the first source it could be established that they were still entangled. The input beam represented a good approximation to an unknown state and the preservation of entanglement showed that quantum features of the state could be successfully transferred. 
%Figure. Figure 4 from BOW03
%\bef
%\vspace{50mm}
%%\leavevmode \epsfbox{Ou92Fig1.eps}
%\caption{\it T-V graph of results from the Bowen et al experiment
%\cite{BOW03}.}
%\label{fig:bowres2}
%\eef
%

\section{Quantum Computation}
\label {sec:QC} 

We have now examined a number of quantum information tasks that have been achieved using optics. We have seen that with some encodings arbitrary control of single qubits can be achieved and specific entangled states can be produced and used as resources for small scale operations. But
what about the more challenging task of quantum computation? The skills so far discussed are insufficient to implement quantum computation. It turns
out that to be able to implement arbitrary processing of information
encoded on a set of qubits it is sufficient to possess at least one
non-trivial two qubit operation, in addition to arbitrary operations on single qubits. 
% Give explicit formula (Nathan)

\begin{figure}[htb]
\begin{center}
\includegraphics*[width=8cm]{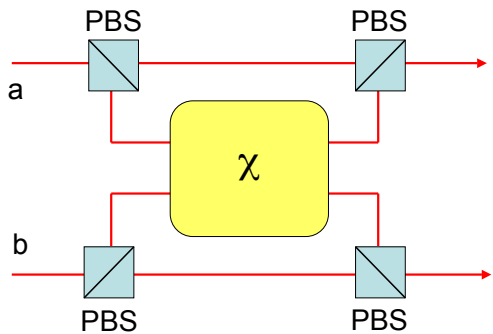}
\caption{Schematic of the implementation of an optical CS gate using a strong cross-Kerr non-linearity $\chi$. PBS are polarizing beamsplitters.}
\label{figA}
\end{center}
\end{figure}

An example of a non-trivial two-qubit gate is the CNOT gate.
In terms
of polarisation qubits its operation is summarised by the following
truth table
\bea
\ket{H}_{c} \ket{H}_{t} & \to & \ket{H}_{c} \ket{H}_{t} \nnum
\ket{H}_{c} \ket{V}_{t} & \to & \ket{H}_{c} \ket{V}_{t} \nnum
\ket{V}_{c} \ket{H}_{t} & \to & \ket{V}_{c} \ket{V}_{t} \nnum
\ket{V}_{c} \ket{V}_{t} & \to & \ket{V}_{c} \ket{H}_{t}
\label{eq:CNOT}
\eea
When the control qubit is in the horizontal state, $\ket{H}_{c}$, the
value of the target qubit $\ket{H}_{t}$ or $\ket{V}_{t}$ is unchanged.
However, when the  control is vertical, $\ket{V}_{c}$, the value of the
target qubit is flipped, horizontal to vertical and vice versa.
The effect of a CNOT gate on superposition states is simply a
superposition of the transformations of equation (\ref{eq:CNOT}). For
example if the control is in the diagonal basis we get the following
transformations
\bea
(\ket{H}_{c} + \ket{V}_{c}) \ket{H}_{t} & \to &
(\ket{H}_{c} \ket{H}_{t} + \ket{V}_{c} \ket{V}_{t}) \nnum
(\ket{H}_{c} + \ket{V}_{c}) \ket{V}_{t} & \to &
(\ket{H}_{c} \ket{V}_{t} + \ket{V}_{c} \ket{H}_{t}) \nnum
(\ket{H}_{c} - \ket{V}_{c}) \ket{H}_{t} & \to &
(\ket{H}_{c} \ket{H}_{t} - \ket{V}_{c} \ket{V}_{t}) \nnum
(\ket{H}_{c} - \ket{V}_{c}) \ket{V}_{t} & \to &
(\ket{H}_{c} \ket{V}_{t} - \ket{V}_{c} \ket{H}_{t})
\label{eq:CNOT2}
\eea
Notice that the resulting output states are the four Bell-states, see
section (\ref{sec:TELEPORTATION}). If we run this interaction back-wards, that is input the Bell states, we see that orthogonal, separable states are outputted, hence enabling efficient Bell-state analysis. Thus the CNOT gate is a very useful device even for small-scale applications. So how might such an interaction
between two photons be implemented? One solution is to use a $\chi_3$ non-linear medium to induce a cross-Kerr effect between two photon modes, as first suggested by Milburn \cite{MIL89}. Ideally the cross-Kerr effect will produce the unitary evolution $\op U_K = \exp[i \chi \op a^{\dagger} \op a \op b^{\dagger} \op b]$, where $\op a$ represents one optical mode and $\op b$ another. Consider the schematic set-up of Fig.\ref{figA}. Two polarization encoded qubits are converted into spatial dual rail qubits using polarizing beamsplitters. One mode from each of the qubits is sent through the cross-Kerr material. The operation of this device on an arbitrary two qubit input state is given by the following evolution:
\bea
\ket{\psi} &\to& \op U_K \ket{\psi} \nonumber\\
&=& e^{i \chi \op{a}_2^\dagger \op{a}_2 \op{b}_1^\dagger \op{b}_1}( \alpha \ket{01}_a\ket{01}_b + \beta \ket{10}_a\ket{10}_b  \nonumber\\
&& \;\;\;\;\;\;\;\;\;\;\;\;\;\;\;\;\;\;\;\;\;\;\;\;\;+ \gamma \ket{10}_a\ket{01}_b  + \delta \ket{01}_a\ket{10}_b ) \nonumber\\
&=&  \alpha \ket{01}_a\ket{01}_b  + \beta \ket{10}_a\ket{10}_b  \nonumber\\
&& \;\;\;\;\;\;\;\;\;\;\;\;\;\;\;\;\  + \gamma \ket{10}_a\ket{01}_b  + e^{i \chi} \delta \ket{01}_a\ket{10}_b  \nonumber\\
&&
\label{c-K}
\eea
Only when both modes entering the Kerr material are occupied is a phase shift induced. If we now choose the strength of the non-linearity such that $\chi = \pi$, the effect is to flip the sign of one element of the superposition. This is called a controlled-sign (CS) gate. If Hadamard gates are placed on qubit $b$, before and after the CS gate (as could be implemented with wave plates, see section \ref{subsec:DRE}) then CNOT operation is achieved with qubit $a$ as the control and qubit $b$ as the target. 

The problem with this idea in practice is that typical non-linear materials have values of $\chi$ that are an order of magnitude of orders of magnitude too small. One might consider making the interaction region of the material very long in order to boost the non-linearity, but such a strategy generally leads to very high levels of loss, which negate the desired effect. Non-linearities close to those required can be realized in cavity quantum electro-dynamic (QED) situations featuring single atoms in cavities of extremely high finesse and small volume \cite{BRU96,TUR95}. This occurs in the so-called {\it strong coupling} regime, in which the dipole coupling between the cavity field and the atom is significantly greater than the relaxation rates of both the cavity and the dipole. Many problems exist with this approach including: the difficulty of coupling photons efficiently into and out of the cavity mode; the need to isolate the cross-Kerr non-linearity from other non-linear effects and; the difficulty in maintaining a constant coupling strength between the atom and the field. A number of ingenious solutions have been suggested \cite{DUA04,NEM04} but remain unproven experimentally to date.

These problems led most to conclude that large scale quantum processing with optics was untenable. However a number of results in the late nineties and early noughties, culminating in the 2001 paper by Knill, Laflamme and Milburn (KLM) \cite{KNI01} led many to change their view. KLM
found a way to circumvent the
problem of needing a huge non-linearity and showed that it was possible to implement efficient quantum computation using only passive
linear optics, photodetectors, and single photon sources. In the following we will first describe how Grover's quantum algorithm can be implemented in a straight forward manner using linear optics. We then describe KLM's more ambitious scheme for general quantum computation and the experimental steps that have so far been taken.

\subsection{Grover's Algorithm}
\label {sec:GA}

\begin{figure}[htb]
\begin{center}
\includegraphics*[width=9cm]{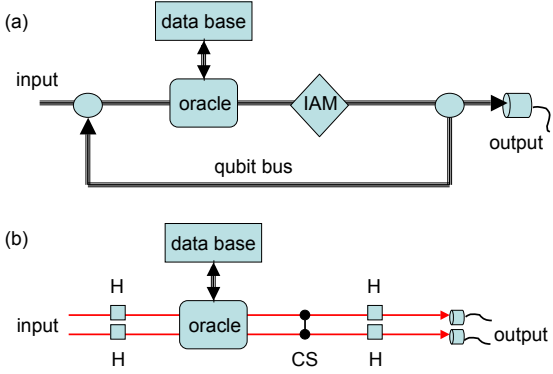}
\caption{Schematic representations of the implementation of Grover's algorithm. (a) General case showing basic flow structure. IAM stands for inversion about the mean. The iterative step is carried out of order $\sqrt{N}$ times, where $N$ is the size of the unstructured data base. (b) Specific implementation for case of $N=4$. The qubits are input in the state $\ket{\bf 0 \bf 0}$ and the output is measured in the computational basis. The 4 possible separable output states unambiguously identify the 4 possible tagged elements with a single query. CS indicates a CS gate and H indicates Hadamard gates.}
\label{figB}
\end{center}
\end{figure}

{\it Grover's algorithm} \cite{GRO97} is an important algorithm in quantum computation 
giving a provable speed-up over classical algorithms in searching an unstructured data-base. The best classical algorithm for finding a single marked item in an unstructured data base is to simply randomly sample. On average, it will take $N/2$ attempts to find the item. Quantum information allows a better solution which is depicted in Fig.~\ref{figB}(a). Dependent on the logical state of a qubit bus a quantum oracle samples the data base. If the nominated item is found, the qubit bus is marked by a phase flip, otherwise the qubit bus is left unchanged. By placing the qubit bus in an equal superposition of all logical states all the data base states can be interogated by a single oracle call. The result is an equal superposition output state with a single phase flip against the qubit state corresponding to the marked item. Inversion about the mean is then performed on the qubit bus, which has the effect of amplifying the marked item with respect to the others. After iterating this process for order $\sqrt{N}$ times the item can found with high accuracy by a measurement on the qubit bus in the computational basis. Although not an exponential speed-up, the improvement in search speed can be quite significant for large $N$. In Fig.~\ref{figB}(b) we show explicitly the smallest non-trivial example: a 4 element data base. In this case a single iteration is sufficient to determine the marked element with unit probability.

We will now consider the implementation of Grovers algorithm in optics. The oracle plays a key role in search algorithms so we shall begin by describing  how, in general, a search oracle can be implemented in optics. Initially we will assume that CNOT gates operating on the principle of Eq.\ref{c-K} are available. We will additionally use the feature that the action of the gate on an unoccupied (vacuum) target mode leaves both modes unchanged, that is
\begin{equation}
\alpha_{1} \ket{H}\ket{0}+\alpha_{2} \ket{V}\ket{0} \to
\alpha_{1} \ket{H}\ket{0}+\alpha_{2} \ket{V}\ket{0}
\label{eq2}
\end{equation}
where $0$ represents the vacuum. We will then show the rather surprising result that such gates are not in fact required in order to implement Grover's algorithm.

\subsubsection{The Oracle}

We require that the oracle, when queried 
by some $n$ qubit state 
will query the corresponding element of an 
unstructured $2^{n}=N$ element classical data-base and either return the same 
$n$ qubit state 
(if the element is not tagged) or a phase flipped version 
(if the element is tagged). We assume the simplest case that only one 
element is tagged. In order that superpositions of the input state 
should be preserved it is clear that we need a classical data-base which 
can be interogated by a quantum particle. Optics provides a 
straightforward solution to this problem.

\begin{figure}
\includegraphics*[width=8cm]{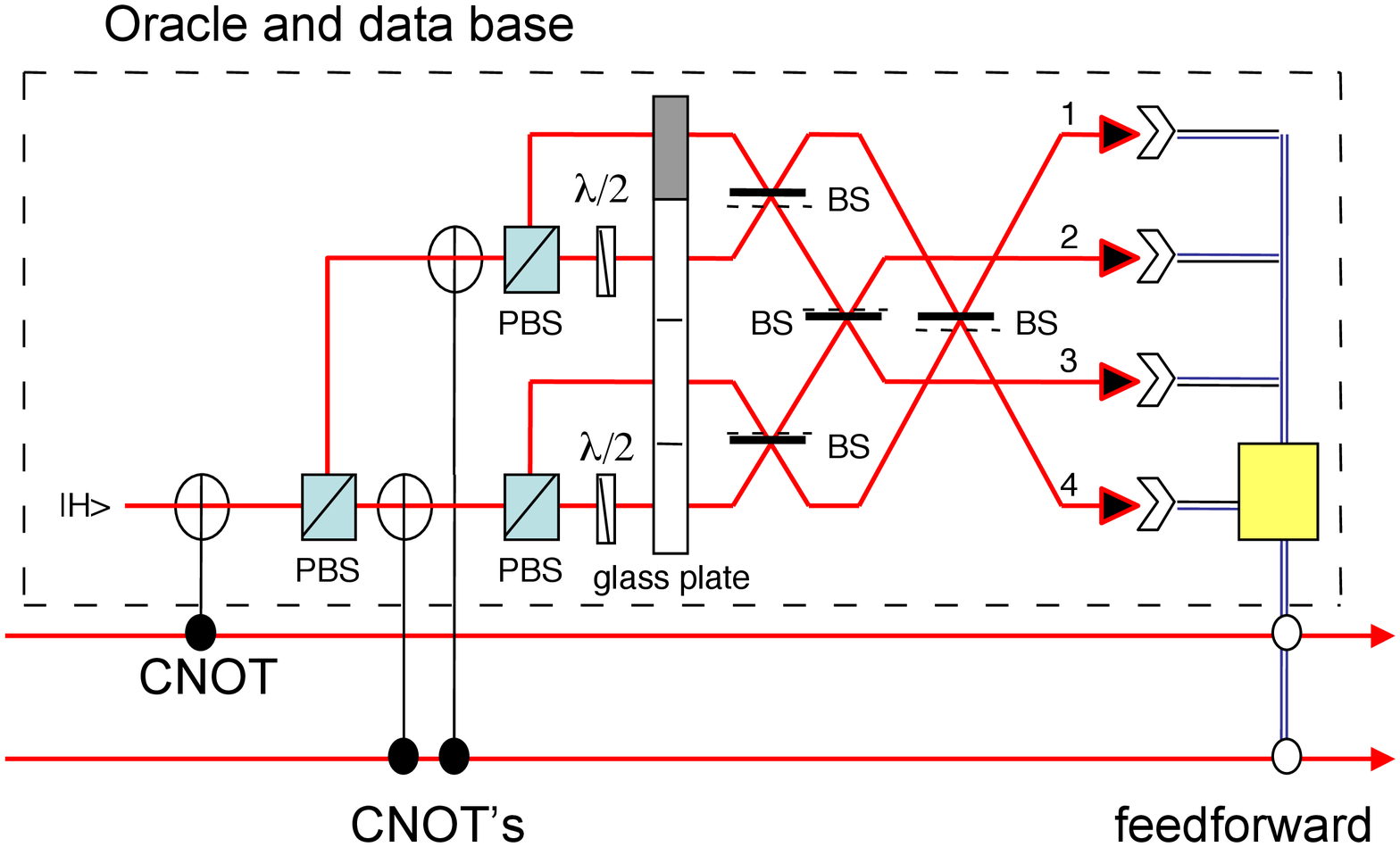}
\caption{Schematic of oracle for 4-bit data base. PBS are polarising 
beamsplitters whilst BS are 50:50 non-polarising beamsplitters. The 
dashed line on the beamsplitters indicates the surface from which 
reflection induces a sign change. Polarising beamsplitters are assumed 
to reflect the horizontal component and transmit the vertical 
component of the incident light. Half-wave-plates are indicated by 
$\lambda/2$ and are oriented at 45 degrees to horizontal, thus 
flipping the polarization of the incident beam. 
Single lines indicate optical rails 
whilst double lines indicate electrical rails. See text for 
description of operation.}
\label{figC}
\end{figure}

To illustrate the technique we shall begin by considering an oracle querying
 a 4 element data-base. We will 
then generalize the result. The proposed arrangement is shown in 
Fig.\ref{figC}. The classical data-base is comprised of
a piece of glass partioned into four domains. One of 
the domains (representing the tagged element) has an optical 
pathlength which is $\lambda/2$ longer than the pathlengths of the 
other domains. Here $\lambda$ is the optical wavelength. 

The principle of the oracle is to direct a single photon through one 
of the four domains as a function of the state of a 2 qubit input. If 
it traverses the tagged domain it will pick up a $\pi$ phase shift 
relative to passage through the other domains. Path information 
carried by the single photon is then erased via a measurement 
protocol leaving the phase flip on the corresponding 2 qubit state.
Because the photon is a quantum particle it can be placed into a 
superposition of traversing various domains simultaneously. 

We start with the qubits in an arbitrary superposition state and the 
``oracle photon'' in the horizontal state:
\begin{eqnarray}
\ket{H}(\alpha_{1} \ket{H}\ket{H}+\alpha_{2} \ket{H}\ket{V}+
\alpha_{3} \ket{V}\ket{H}+\alpha_{4} \ket{V}\ket{V})
\label{eq3}
\end{eqnarray}
Here the ordering of the kets from left to right in Eq.\ref{eq3} 
corresponds to the rail sequence from top to bottom in 
Fig.\ref{figC}.
Next a CNOT gate with the first qubit as the control and the oracle photon 
as the target is applied. Depending on the value of the first qubit the 
polarization of the photon is either left alone or flipped to $V$. The 
oracle photon then passes through a polarizing beamsplitter which 
separates the polarization modes into two separate spatial paths. We 
now apply CNOT's with the second qubit as their control and each of 
the new oracle photon spatial modes as targets. Subsequent polarizing 
beamsplitters again divide polarization modes into different spatial 
modes resulting in four different paths that the photon can take. Each of the 
different paths is uniquely determined by one of the four possible 
logical basis input states of the qubit. Thus the state of the system 
has now evolved to:
\begin{eqnarray}
\alpha_{1} \ket{p_{1}}\ket{H}\ket{H}&+&\alpha_{2} \ket{p_{2}}\ket{H}\ket{V}\nonumber\\
&+&
\alpha_{3} \ket{p_{4}}\ket{V}\ket{H}+\alpha_{4} \ket{p_{3}}\ket{V}\ket{V} 
\end{eqnarray}
where we have used the $\ket{p_{1}} \equiv \ket{1,0,0,0}$, 
$\ket{p_{2}} \equiv \ket{0,1,0,0}$, etc. 
The oracle photon then passes through the glass plate data base. For 
concreteness we will assume that it is the first domain which has the 
increased path length, thus the state of the system after the 
interaction with the data base is
\begin{eqnarray}
-\alpha_{1} \ket{p_{1}}\ket{H}\ket{H}&+&\alpha_{2} \ket{p_{2}}\ket{H}\ket{V}\nonumber\\
&+&
\alpha_{3} \ket{p_{4}}\ket{V}\ket{H}+\alpha_{4} \ket{p_{3}}\ket{V}\ket{V}  
\end{eqnarray}
The tag has successfully been attached to the qubit state however the 
oracle photon still carries information about the qubit state which 
must be erased. This could be done by reversing the sequence of gates 
used before the data-base as was suggested by Nielsen and Chuang \cite{NIE00}. However, the need 
for more quantum gates can be avoided by employing a measurement based erasure protocol. 
This is achieved by mixing all the possible photon paths on 50:50 
beamsplitters. There is then an equal probability of finding the 
photon in any of the four paths, thus erasing the qubit information. 
The photon is then detected. Depending on where the photon is 
found the following phase corrections must be made to the qubits:
(i) If the photon is counted at detector 1, do nothing; (ii) if the 
photon is counted at the second detector then a phase shift 
(defined by $Z\ket{H}=\ket{H}$, $Z\ket{V}=-\ket{V}$ and implementable 
with a quarter-wave plate) is applied to qubit 2; (iii) if the 
photon is counted at the third detector then a phase shift is applied 
to both qubits and; (iv) if the 
photon is counted at the fourth detector then a phase shift is applied 
to the first qubit. After the correction has been made the qubits are 
left (up to a global phase factor) in the state
\begin{eqnarray}
-\alpha_{1} \ket{H}\ket{H}+\alpha_{2} \ket{H}\ket{V}+
\alpha_{3} \ket{V}\ket{H}+\alpha_{4} \ket{V}\ket{V} 
\end{eqnarray}
which is the required state.

This scheme is easily generalized to larger data-bases. For example 
to search an eight element data-base we require a three qubit register 
and a glass plate with eight domains. The first two steps of the 
protocol run the same as before with the photon being fanned out to 
four different paths. Now four CNOT's, all controlled by the third 
qubit and with the four photon paths as their respective targets, 
direct the oracle photon into eight possible paths, each uniquely determined 
by the qubit values. The paths are passed through the data-base, 
resulting in tagging, and the
qubit information is erased by mode mixing followed by detection in an 
analogous way to the four element protocol. 
It is clear that the scheme can be further 
expanded in this way to any finite sized data-base. In general the oracle 
will require $(N-1)$ CNOT gates where $N=2^{n}$ is the data-base size 
and $n$ is the number of qubits in the register. 

\subsubsection{Grover's Algorithm with Linear Optics}

In the previous section we showed how a search oracle could in general be implemented in an optical quantum computation circuit, given a cross-Kerr type two-qubit gate. We now consider the specific case of Grover's algorithm and show that in fact the entire algorithm can be performed using only linear optics.

Consider the initial interaction between the qubit bus and the oracle. In Grover's this step is used to instruct the oracle to make an equal superposition query of all data base elements. However, this can be achieved using only linear optics by simply fanning out a single photon mode into $N$ modes using 50:50 beamsplitters. After the oracle the result is read to the qubit bus for the processing step of inversion about the mean. After processing it is read back to the oracle that then queries the data base again. But actually it is unnecessary to read the information back to a qubit bus in order to perform the processing. As shown by Kwiat and Zeilinger \cite{REC94}, any unitary operation can be performed on a unary data bus using only linear optics. A unary data bus is one in which a particular number, $n$, is represented by having the $nth$ bit flipped with respect to all the others. For the optical bus this is represented in single rail logic (see section \ref{subsec:SRE}) by having only the $nth$ mode occupied by a single photon. This is in contrast to a binary data bus in which $n$ would be represented by the binary digit $n_{base 2}$. 

In a general quantum computation circuit such a unary encoding would lead to an exponential expense in the number qubits needed in comparison to a binary encoding and would in most cases quickly nullify any gain made through the quantum approach. However, for the specific case of a search algorithm like Grover's, it is a required step to introduce a unary qubit bus in order that the unstructured data base can be queried. It is thus of no benefit to continually shift  back and forth between the binary and unary qubit buses and is just as efficient to remain in the unary qubit bus and perform all the processing using linear optics.

A number of groups have performed in principle demonstrations of Grover's algorithm using linear optics. Kwiat el al \cite{KWI99} searched a 4 element data base. They used a combination of polarization and spatial encoding to form the unsorted data bus and a Sagnac interferometer to form a passively stable interferometric arrangement. The data base was electro-optically "programmed" via waveplates and a Pockell cell. They achieved around 90\% probability of successfully identifying the marked data base element. Bahattacharya et al \cite{BAH02}, motivated by a proposal of Lloyd \cite{LLO00}, were able to search up to a ~32 element data base, represented by a thin groove in  glass plate. They used a standing-wave cavity to achieve repeated interactions with the data base and Fourier optics to produce the required inversions about the mean. 

In neither of these experiments were single photon states used. Rather, bright optical pulses containing huge numbers of photons passed through the systems. Although in the case of the Kwiat experiment it would have been reasonably straightforward to run the experiment with single photons \cite{NOTE1}, a significantly more difficult set-up would be required to allow the Bahattacharya experiment to be run with any reasonable efficiency in the single photon domain. This prompts the question of whether, if the basic effect of Grover's algorithm is observable with bright beams of light, the algorithm should be considered "quantum". We now briefly address this question. 

\subsubsection{Is Grover's Algorithm Quantum?}

In order to answer the question "Is Grover's Algorithm Quantum?" we first need to define what we mean by "a single query of the data base" and what we mean by "quantum". We will adopt the following definitions:
\begin{enumerate}
\item A single query of the data base occurs when, on average, a single photon interacts with the data base.
\item The algorithm will be considered quantum if it is necessary for entanglement to exist between the optical modes comprising the unsorted data bus in order to achieve the $\sqrt{N}$ scaling.
\end{enumerate} 
Notice that by this definition neither of the experiments so far performed strictly realized Grover's algorithm as both involved many photons interacting with the oracle per query. We now consider three examples that satisfy the query definition, two of which involve entanglement and one of which does not.

Firstly, the implementation of the oracle represented in Fig.~\ref{figC} clearly satisfies the query definition and also produces a $\sqrt{N}$ scaling. The state of the data bus just before the first interaction with the data base is
\beq
\ket \psi = \ket{100...00} + \ket{010...00} + .....\ket{000...01} 
\label{grovst}
\eeq
which clearly exhibits modal entanglement.

Can we remove the entanglement and still retain the $\sqrt{N}$ scaling. As a second example we could consider using a weak coherent state with amplitude $\alpha = 1$ as the input instead of a single photon state. We still satisfy the query requirement on average and the algorithm still achieves a $\sqrt{N}$ scaling. Because we are using a weak coherent state there is a $37\%$ probability for any particular query that we will inject vacuum into the circuit and hence get a null result. However this efficiency is constant, independent of the size of the data base and thus does not affect the scaling. Is there still entanglement present? At first sight the answer to this question may appear to be "no", as the unitary evolution of a coherent state through a beamsplitter does not produce entanglement. However the photon counters used to detect the final state have microscopic resolution and reject the vacuum state realizations. As a result, on the occasions the algorithm succeeds, the detectors post-select circuit states similar to that in Eq.\ref{grovst}. Again, entanglement is seen to be present.

To avoid both unitary and post-selected entanglement we consider a third example in which again we use a weak coherent state with amplitude $\alpha = 1$ as the input but now use homodyne detection of the final state. Homodyne detection measures the field amplitude and so does not resolve individual quanta. As a result it cannot post-select entanglement from coherent state inputs. Now the state of the data bus just before the first interaction with the data base is
\beq
\ket \psi = \ket{\alpha', \alpha',..., \alpha'} 
\label{cgrovst}
\eeq
where $\alpha' =1/\sqrt{N}$. Eq.\ref{cgrovst} is clearly a separable state. The output state after the correct number of iterations will be approximately
\beq
\ket \psi_{out} = \ket{0, 0,.., \alpha,...,0} 
\label{cgrovout}
\eeq
where the mode position of the displaced state gives the marked element. However, using homodyne detection means that there is vacuum noise associated with the unoccupied modes meaning we cannot unambiguously identify the marked element. Indeed the signal to noise with which we can identify the correct element is now $1/N$. In order to maintain a constant signal to noise, say of $1$, we need to repeat the algorithm $\sqrt{N}$ times. As a result we find that the number of data base queries scales with $N$ in this case, just as for the classical algorithm. Given our energy constraint it is hard to imagine how this problem can be avoided whilst still maintaining separable states.

These examples suggest strongly that, given our definitions, Grover's algorithm should be considered a quantum algorithm not-with-standing its classical field analogues.

\subsection{Linear Optical Quantum Computation}
\label {sec:LOQC}

As we have already mentioned, it is not possible to construct a general quantum computation network using the unsorted encoding scheme without incurring an exponential overhead. In the following we will describe the scheme of KLM, in which the standard dual-rail qubit encoding is used, but arbitrary processing is achieved without Kerr type non-linearity or an exponential overhead. Instead the KLM toolbox comprises: single photon sources; photon counting detectors and; electro-optic feed-forward.

There are three tiers to the KLM scheme:
\begin{enumerate}
\item Non-deterministic two qubit gates which can be used to produce entangled resource states.
\item Non-deterministic teleportation gates which are driven by entangled resource states and fail by accidentally measuring the value of the qubit.
\item Error correcting codes that protect the qubits from accidental measurement during the application  of the teleportation gates and hence allow scale up of universal circuits without an exponential overhead.
\end{enumerate} 
We now discuss each of these tiers in turn and the experimental progress that has been made towards quantum computing based on this paradigm.

\subsubsection{Non-Deterministic Entangling Gates}
\label{sec:NDEG}

\begin{figure}
\includegraphics*[width=9cm]{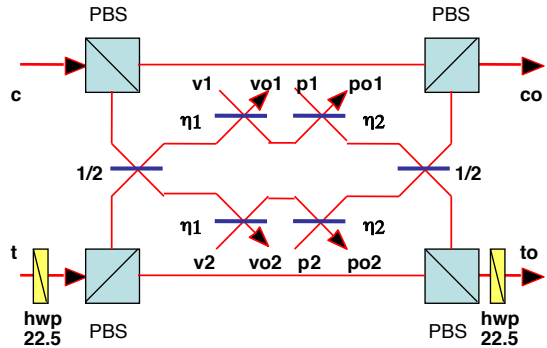}
\caption{Schematic representation of a non-deterministic CNOT gate.
Polarization encoded qubits are injected at $c$ and $t$. Acillary photons are injected at $p1$ and $p2$. Successful operation is heralded by the detection of no photons at
outputs $vo1$ and $vo2$ and the detection of one and only one photon
at each of outputs $po1$ and $po2$. PBS are polarizing beamsplitters and hwp are half-wave plates.}
\label{figD}
\end{figure}

At the first level, KLM introduced two qubit gates that could take separable, single photon inputs, and produce entangled outputs. In particular KLM showed how to make a CNOT gate that was
non-deterministic, but heralded. That is, the gate does not always
work, but an independent signal heralds successful operation. A
somewhat simplified version of this gate is shown in Fig.(\ref{figD}) \cite{RAL01}.
In addition to the single photon, polarisation
qubits incident at ports $c$ (control) and $t$ (target), the gate also
has ancilla inputs comprising
two vacuum input ports, $v1$ and $v2$, and two single photon input
ports, $p1$ and $p2$. The beamsplitter reflectivities are given by
$\eta_{1}=5-3 \sqrt{2}$ and $\eta_{2}=(3-\sqrt{2})/7$. It can be shown
that when no photons are detected at outputs $vo1$ and $vo2$, and one
and only one photon is detected at each of $po1$ and $po2$, then the gate
has succeeded and the photon qubits exiting through $co$ and $to$ have
had the CNOT transformation applied to them. The probability of
successful operation is $\eta_{2}^{2} \approx 0.05$. Recently it has been proved by Eisert \cite{EIS05} that $1/16$ is the upper bound for success probability for a gate of this type (as achieved by the original KLM proposal).

Even at this first level the technical requirements
are demanding. Four photons need to simultaneously enter the circuit.
The detections at $po1$ and $po2$ have to distinguish between zero,
one or two photons. Any inefficiency in the production or detection of
photons will lead to mistakes and rapidly erase the operation of the
gate. High visibility single photon and two photon (HOM type) interference
are required simultaneously: as a result excellent mode-matching and
photon indistinguishability are essential. 

\begin{figure}
\includegraphics*[width=9cm]{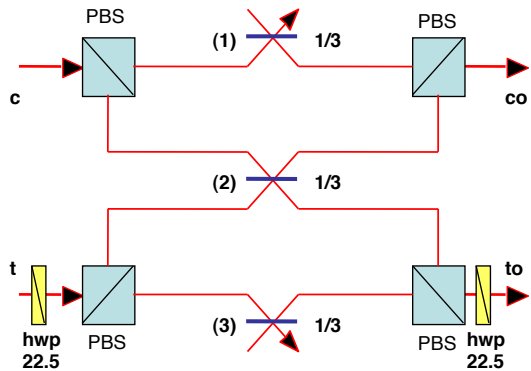}
\caption{Schematic representation of non-deterministic coincidence CNOT gate.
Polarization encoded qubits are injected at $c$ and $t$. PBS are polarizing beamsplitters and hwp are half-wave plates.}
\label{figE}
\end{figure}

A significantly simpler CNOT design can be realised by
working in coincidence as discussed by Ralph et al \cite{RAL02} and independently by Hofmann and Takeuchi \cite{HOF02}.
In particular we can allow the photon
qubits to be their own ancilla, such that only two photons are required.
The gate is shown schematically in Fig.(\ref{figE}). Consider
the input $\ket{H}_{c} \ket{H}_{t}$. The target waveplate produces
the transformation
\beq
\ket{H}_{c} \ket{H}_{t} \to {{1}\over{\sqrt{2}}}
\ket{H}_{c} (\ket{H}_{t} + \ket{V}_{t}) \nnum
\eeq
The polarizing beamsplitters then spatially separate the polarization
modes of the two beams. An array of different possibilities are
present after the middle beamsplitters, however, we select (by
postselection) only those where a photon arrives at both the target
and control outputs. There are two ways for this to happen: the
control photon must take the top path and reflect off beamsplitter 1;
the target photon may take its upper path and reflect off beamsplitter 2 or
take the bottom path and reflect off beamsplitter 3. In both cases the
effect is just to reduce the amplitude of the successful components
by a factor of $1/3$. The output state is then transformed by the
second target waveplate such that
\beq
{{1}\over{3}} {{1}\over{\sqrt{2}}}
\ket{H}_{c} (\ket{H}_{t} + \ket{V}_{t}) \to {{1}\over{3}}
\ket{H}_{c} \ket{H}_{t} \nnum
\eeq
Similarly the input $\ket{H}_{c} \ket{V}_{t}$ is unchanged by passage
through the circuit, other than a $1/3$ reduction in amplitude.

Things are different when the control is in the vertical state.
Consider the input state $\ket{V}_{c} \ket{H}_{t}$.
The target waveplate produces the transformation
\beq
\ket{V}_{c} \ket{H}_{t} \to {{1}\over{\sqrt{2}}}
\ket{V}_{c} (\ket{H}_{t} + \ket{V}_{t}) \nnum
\eeq
Now there are three ways for a successful detection to occur. The
control takes its lower path. If the
target photon takes the bottom path then both must reflect off their
respective beamsplitters as before, simply reducing the amplitude by
$1/3$. However, if the target photon follows its upper path then there
are two possibilities at beamsplitter 2: either both photons may be
reflected, giving an amplitude of $1/3$, or; both may be transmitted,
giving an amplitude of $-2/3$. If the photons are indistinguishable
then these amplitudes are added giving a total amplitude for that
component of $-1/3$! Thus when the
polarization modes are recombined the state carries a minus sign on
one target component and the second target waveplate makes the
transformation
\beq
{{1}\over{3}} {{1}\over{\sqrt{2}}}
\ket{V}_{c} (\ket{H}_{t} - \ket{V}_{t}) \to {{1}\over{3}}
\ket{V}_{c} \ket{V}_{t} \nnum
\eeq
and the value of the target qubit is flipped as required. Similarly
the circuit does the transformation $\ket{V}_{c} \ket{V}_{t} \to
1/3 \ket{V}_{c} \ket{H}_{t}$. Hence CNOT operation is realized
whenever a coincidence is recorded. The probability of success is
$(1/3)^{2} = 1/9$.

J.O'Brien and G.Pryde et al used this technique to demonstrate
CNOT operation for single photon qubits \cite{OBR03}. 
In their experiment a pair of polarisation beam
displacers was used to create an interferometrically stable
configuration. Down conversion was used to produce suitably pure photon pairs in separable polarization states which were injected into the gate. State tomography
(see section \ref{sec:CHARPQ}) was used to compare the experimental output with the expected outcome. Good agreement was found. In particular entanglement could be produced as expected. In later experiments full process tomography was carried out \cite{OBR04} from which an average fidelity of around 90\% for the gate was calculated.

%%
%%
%%Figure. Figure 1(c) OBR03
%\bef
%\vspace{50mm}
%%\leavevmode \epsfbox{Ou92Fig1.eps}
%\caption{\it Experimental layout of the demonstration of  a CNOT gate 
%by O'Brien
%and Pryde et al. \cite{OBR03}.}
%\label{fig:obrset}
%\eef
%%

%
%The truth table performance ranged from about 95\% correct for ``control-off''
%operation, to 73\% for ``control-on'' operation. The difference is
%attributed to the need for both single and two-photon interference
%when the control is on. All four Bell-states were produced as
%described in equ.(\ref{eq:CNOT2}) with fidelities ranging from 77\% 
%to 87\%, thus
%demonstrating the ability of the gate to produce entanglement.
%Experimental density matrices for two of the Bell-states are shown
%in Fig.(\ref{fig:obrres}).
%\index{state!Bell}\index{Bell!state}

%
%%
%%
%%Figure. Figure 4 OBR03
%\bef
%\vspace{50mm}
%%\leavevmode \epsfbox{Ou92Fig1.eps}
%\caption{\it Density matrices for two of the Bell states produced by
%the CNOT gate of O'Brien and Pryde et al.\cite{OBR03}: (a) real part
%of density matrix for $\ket{\psi-}$ Bell state (imaginary part zero)
%and real and imaginary parts of coresponding
%measured density matrix; (b)  real part
%of density matrix for $\ket{\phi+}$ Bell state (imaginary part zero)
%and real and imaginary parts of corresponding measured density matrix }
%\label{fig:obrres}
%\eef

%
%

\subsubsection {Teleportation Gates}

We now proceed to the second tier of the KLM scheme. Although the gates discussed in the previous section give us access to non-trivial two-qubit operations and small scale circuits, they are ultimately not scaleable. A cascaded sequence of
such non-deterministic gates would be useless for quantum computation because
the probability of many gates working in sequence decreases
exponentially. In order to make a scaleable system we must move to {\it teleportation gates}. 
%This
%problem may be avoided by using teleportation. 

\begin{figure}
\begin{center}
\includegraphics*[width=10cm]{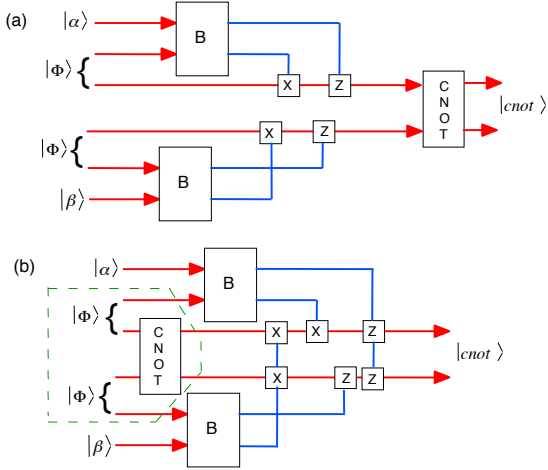}
\caption{ Schematic representation of gate operation via
teleportation. Figures (a) and (b) are equivalent, yet in (b) a
non-deterministic CNOT gate is sufficient as failure only destroys the
entanglement: the operation can be repeated till successful without
losing the qubit.}
\label{figF}
\end{center}
\end{figure}

The
idea that teleportation can be used for universal quantum computation was
first proposed by Gottesman and Chuang \cite{GOT99}.
Consider the quantum
circuit shown in Fig.(\ref{figF}) (a). Two unknown qubits are individually
teleported and then a CNOT gate is implemented. Obviously, but
not very usefully, the result is CNOT operation between the input and
output qubits. However, the commutation
relations between CNOT and the $X$ and $Z$ operations used in the
teleportation are quite
simple, such that in the circuits of Fig.\ref{figF} the 
alternatives (a) and
(b) are in fact
equivalent. But in the circuit of Fig.\ref{figF} (b)
the problem of implementing a
CNOT gate has been reduced to
that of producing the required entanglement resource. The main point
is that this need not be done
deterministically. Non-deterministic CNOT gates could be used in
a trial and error manner to build up the necessary resource off-line. From this point of view the gates of the previous section can be regarded as entanglement factories - producing entangled states for use in teleportation. Alternatively we can note that some photon sources, such as parametric down-conversion, can produce entangled photons directly.
%Figure. New figure to be supplied
%
%
%\bef
%\vspace{50mm}
%%\leavevmode \epsfbox{Ou92Fig1.eps}
%\caption{\it Schematic representation of gate operation via
%teleportation. Figures (a) and (b) are equivalent, yet in (b) a
%non-deterministic CNOT gate is sufficient as failure only destroys the
%entanglement: the operation can be repeated till successful without
%losing the qubit.}
%\label{fig:CNOT2}
%\eef
%

\begin{figure}
\begin{center}
\includegraphics*[width=9cm]{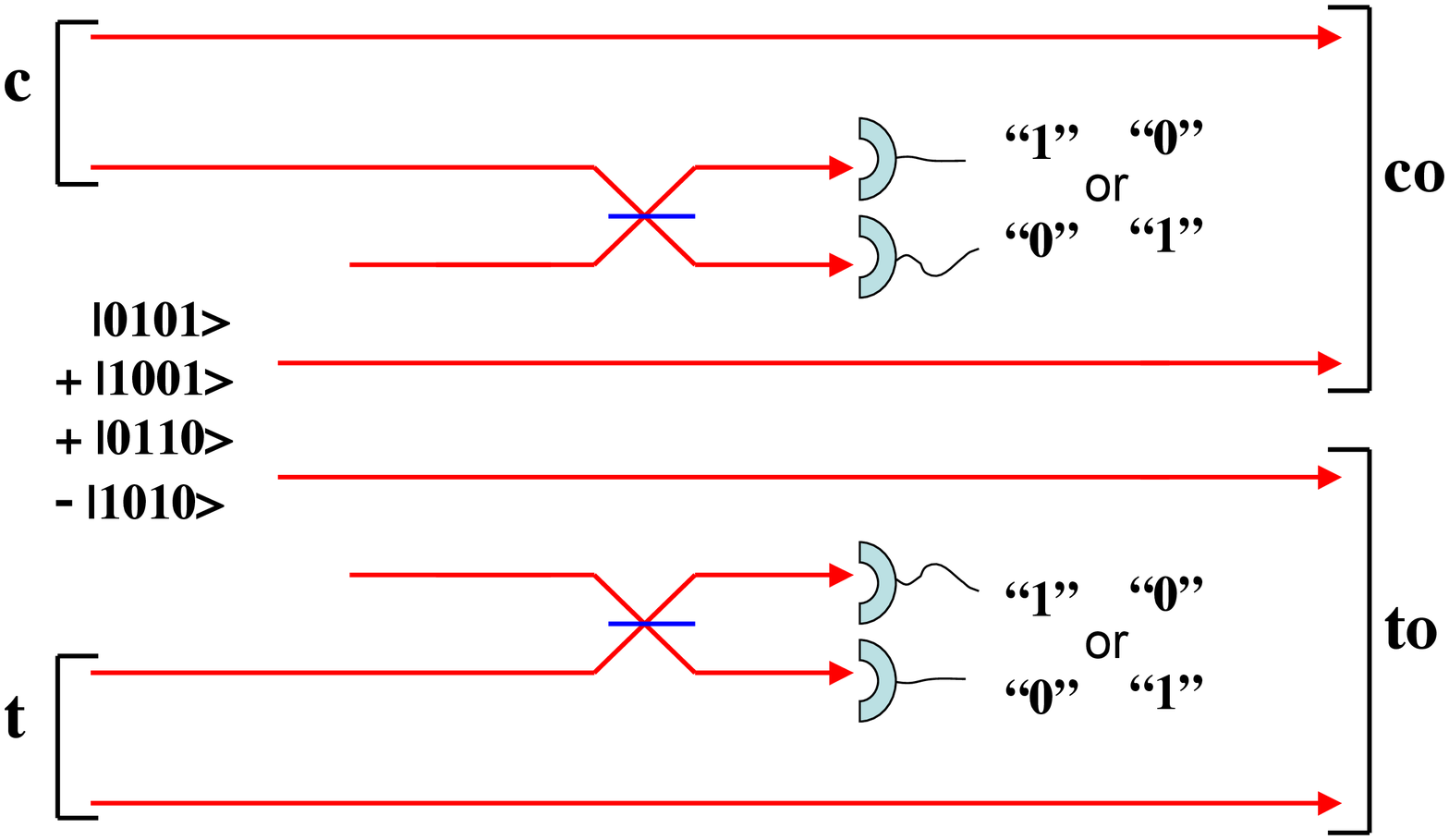}
\caption{ Schematic representation of optical CS gate operation via
teleportation. Success is heralded by a single photon being detected at each of the two pairs of detectors. If zero (two) photons are detected at one of the detector pairs then the corresponding qubit has been measured to be in the zero (one) logical state and the gate has failed. The probability of success of the gate is 25\%.}
\label{figG}
\end{center}
\end{figure}

The simplest teleportation gate is shown in Fig.\ref{figG}. The heart of the gate is a teleported single-rail CS gate. CS operation on single-rail qubits can equally well be used to produce CS operation on dual-rail qubits simply by adding additional rails which do not participate in the interaction (Fig.\ref{figG}). The entangled resource is the state 
\beq
\ket{0101} + \ket{0110} + \ket{1001} - \ket{1010}
\eeq
which can be interpreted as two single rail Bell states which have had a CS gate applied between them in analogy with the resource state in Fig.\ref{figF} (b). Alternatively one can recognize this state as the dual-rail Bell state $\ket{0101} + \ket{1010}$ with a Hadamard gate applied to the second qubit. Such a state can be generated directly by down conversion.  This latter interpretation is due to Pittman, Jacobs and Franson \cite{PIT01}. 

Single-rail partial Bell measurements are used, as described in section \ref{sec:TELEPORTATION}. These fail 50\% of the time thus the probability of success of this gate is 25\%. Because the partial Bell measurements fail by measuring their inputs in the computational basis, so the teleportation gate fails in the same way by measuring the logical values of the input qubits. The key to scale up is that this failure mode can be encoded against, as will be described in the next section.

An in principle demonstration of this gate was made by Gasparoni et al \cite{GAS04}. A femto-second pump pulse was double passed through a down conversion crystal to produce two entangled pairs of photons, one of which was used as the entangled resource whilst the other pair served as the input qubits. Gate operation was demonstrated with an average gate fidelity that can be estimated to be about 78\%. Other demonstrations of this type of gate have been made by Pittman {\it et al} \cite{PIT03} and Zhao {\it et al} \cite{ZHA04}. Although this gate is in principle heralded it is important to note that the current low efficiencies of the sources and detectors mean all experiments so far have relied on coincidence detection.

KLM showed that by using more complex entangled states teleportation gates with a higher probability of success could be implemented. Consider the entangled state
\beq
\ket{0011} + \ket{1010} + \ket{1100} .
\label{T3}
\eeq
If the first two modes of this entangled state are mixed with an input qubit on a beam tritter (i.e. an arrangement of beamsplitters that coherently mixes three input fields in equal ratios) and are then photon counted, teleportation can be achieved for certain measurement outcomes. For example if the measurement result is "$100$", then the conditional state of the remaining two modes of the entanglement will be $\alpha \ket{10} + \beta \ket{11}$. The state has been successfully teleported to the last mode. The other mode is definitely in the "$1$" state and can be discarded. Similarly if the measurement result "$200$" was recorded then the conditional state of the remaining two modes of the entanglement will be $\alpha \ket{00} + \beta \ket{10}$. Now the state has been successfully teleported to the first of the remaining modes, whilst the other mode is definitely in the "$0$" state and can be discarded. Other measurement results will require phase shifts of $\pm \pi/\sqrt{3}$ to recover the qubit. If no photons are counted (assuming unit detection efficiency) or $3$ photons are counted then the original qubit must have been in the "$0$" or "$1$" states respectively and the teleportation fails by measurement of the qubit. The combined probability of these failure modes is $1/3$ so the teleporter has a $2/3$ probability of success.

This more efficient teleporter can then be used to implement a teleportation gate with a probability of success of $4/9$ using the entanglement resource
\bea
& & \ket{00110011} + \ket{00111010} + \ket{00111100} + \ket{10100011} \nonumber \\
&&\;\;\; - \ket{10101010} + \ket{10101100} +\ket{11000011}\nonumber \\
&&\;\;\; \;\;\; + \ket{11001010}+\ket{11001100} .
\eea
which is two entangled states of the form of Eq.\ref{T3} with CS gates applied between all the possible combinations of output modes. This state could be produced non-deterministically using the level one gates but with considerable overhead. Higher order gates with even better probabilities of success are possible with correspondingly more complicated resource entanglement. Because of the high cost of producing the entanglement this is not a viable approach to scalability. Instead, in the next section, we discuss the much more promising approach of error encoding. 

\subsubsection{Error Encoding Against Teleportation Failure}

In the previous section we have seen that teleportation gates can be implemented which have higher probability of success than the first tier non-deterministic gates. A key feature of these gates is that failure results in the measurement of the logical values of the qubits. KLM introduced an error correction code  to protect against such computational basis measurements (Z-measurements) of the qubits. A logical qubit can be encoded across 2 physical qubits as \cite{KNI01}  
\begin{equation}
    \ket{\phi}^{(2)} = \alpha (|\bf 0 \rangle 
    |\bf 0 \rangle + |\bf 1 \rangle |\bf 1 \rangle) + \beta (|\bf 0 \rangle 
    |\bf 1 \rangle + |\bf 1 \rangle |\bf 0 \rangle)
    \label{eq1}
\end{equation}
This is a parity encoding, that is the ``zero'' state is represented 
by an equal superposition of all the even parity combinations of the 2 qubits whilst the ``one'' 
state is represented by all the odd parity combinations. Notice that if a 
Z-measurement is made on either of the physical qubits of the state in 
Eq.\ref{eq1} and the result ``0'' is obtained, then the state 
collapses to an unencoded qubit, however the superposition is 
preserved. Similarly if the measurement result is ``1'' a bit-flipped 
version of the unencoded qubit is the result, but again the 
superposition is preserved so the qubit can be recovered. 

This encoding thus enables recovery from teleportation gate failure and so improves the probability of success of the gate by allowing second attempts. An in principle demonstration of this encoding was made by O'Brien {\it et al} \cite{OBR05} using the two photon CNOT gate discussed in section \ref{sec:NDEG} to produce the required parity encoded states, where the CNOT gate takes an unencoded qubit as its target input and a diagonal state as its control input. It was shown that measurement of either physical qubit led to the expected unencoded qubit being projected onto the remaining photon to an accuracy of greater than 90\% fidelity. 

Notice that a two-qubit (and thus non-deterministic) gate is needed to produce the parity encoding. It is not immediately obvious that producing encoded states non-deterministically which then can be used to improve the performance of more non-deterministic gates, is a winning strategy. KLM showed however, that provided you start with teleporters with a probability of success greater than 50\%, this strategy does improve gates success. For example a $2/3$ teleporter used with the parity encoding leads to a CS gate success probability of about 58\% (as opposed to 44\% without encoding). In order to further improve the probability of success KLM concatenates the two qubit parity code. For example the next level up logical qubit is given by 
\begin{eqnarray}
    \ket{\phi}_{L4} &=& \alpha (|{\bf 0} \rangle^{(2)} 
    |{\bf 0}  \rangle^{(2)}  + |{\bf 1} \rangle^{(2)}  |{\bf 1}  \rangle^{(2)} ) \nonumber\\
    &&\;\;\;\;\;\;\;\;+ \beta (|{\bf 0}  \rangle^{(2)}  
    |{\bf 1}  \rangle^{(2)}  + |{\bf 1}  \rangle^{(2)}  |{\bf 0}  \rangle^{(2)} )
    \label{eq2}
\end{eqnarray}
High probabilities of success are obtained after a few levels of concatenation, leading to the claim of a scalable system. 

\subsubsection{Parity States and Cluster States}

KLM was a major step forward both in opening the door to small-scale demonstrations of optical quantum circuits, and in pointing the way towards a scalable system. However, in its original form the resources required for scale-up were exorbitant. For example the number of Bell pairs needed to implement a single CS gate with 95\% probability of success using the original KLM approach can be estimated to be in the 10,000's. Fortunately, considerable progress has been made in recent years in reducing this overhead \cite{YOR03, NIE04, HAY04} with the most efficient approaches requiring of order 100 Bell pairs for a CS with $> 95\%$ success \cite{BRO05,GIL05}. Two related but distinct approaches have emerged which we now discuss.

An alternative way to scale up the parity states, introduced by Hayes et al \cite{HAY04}, is not to concatenate the code as per Eq.\ref{eq2}, but instead to increase it incrementally. Hence a 
logical qubit can be encoded across $n$ qubits by representing logical 
``zero'' by all the even parity combinations of the $n$ qubits and 
logical ``one'' by all the odd parity combinations. This code retains the feature that if the 
logical qubit is encoded across $n$ physical qubits then a 
computational basis measurement on any one of the qubits reduces the state to a logical 
qubit encoded across $(n-1)$ physical qubits (with the possible need for a bit-flip). 
Specifically, this parity encoding is given by
\begin{eqnarray}
\label{parity}
\ket{{\bf 0}}^{(n)} & \equiv & (\ket{+}^{\otimes n}+\ket{-}^{\otimes
n})/\sqrt{2}\nonumber \\  
\ket{1}^{n} & \equiv & (\ket{+}^{\otimes n}-\ket{-}^{\otimes
n})/\sqrt{2},
\end{eqnarray}
where  $\ket{\pm} = (\ket{{\bf 0}} \pm \ket{{\bf 1}})/\sqrt{2}$.  

There are two operations which are easily performed on parity encoded
states: a rotation by an arbitrary amount around the $x$ axis of the
Bloch sphere (i.e. $X_\theta=\cos(\theta/2) I + i\sin(\theta/2)X$) \cite{NOTE2}, 
which can
be performed by applying that operation to any of the physical qubits and;  a $Z$ operation, which can be performed by applying $Z$ to
\emph{all} the physical qubits (since the odd-parity states will acquire an
overall phase flip).  

%A key operation we will use is the partial Bell state
%measurement \cite{Wei94,Brau95}.  This consists of mixing two physical qubits
%on a polarising beam splitter followed by measurement in the
%diagonal-antidiagonal basis.  A successful event occurs when a photon is
%counted at each out put of the beamsplitter. An unsuccessful event occurs when
%both photons appear at one of the outputs. When successful it projects onto the
%Bell states $\ket{00}+\ket{11}$ and $\ket{00}-\ket{11}$. When unsuccessful it
%projects onto the separable states $\ket{01}$ and $\ket{10}$, thus measuring
%the qubits in the computational basis. 
The teleportation gates are reduced to just partial single-rail and dual-rail Bell-state measurements. A dual-rail Bell measurement can be used to add $n$ physical qubits to a parity encoded state using a resource of $\ket{{\bf 0}}^{(n+2)}$. This is referred to as type-II fusion ($f_{II}$)  \cite{BRO05}. The result of $f_{II}$ is
\begin{equation}
	f_{II}\ket{\psi}^{(m)}\ket{{\bf 0}}^{(n+2)}\rightarrow\left\{ \begin{array}{cl}
		\ket{\psi}^{(m+n)} & \mbox{(success)}\\
		\ket{\psi}^{(m-1)}\ket{{\bf 0}}^{(n+1)} & \mbox{(failure)}
	\end{array}\right.
	\label{encoding}
\end{equation}
When successful (with probability $1/2$), the length of the parity qubit is
extended by $n$. A phase flip correction may be necessary depending on the
outcome of the Bell-measurement.  If unsuccessful a physical qubit is removed
from the parity encoded state, and the resource state is left in the state
$\ket{{\bf 0}}^{(n+1)}$ (which may be recycled).  This encoding procedure is equivalent
to a gambling game where we either lose one level of encoding, or gain $n$
depending on the toss of a coin. Clearly if $n \ge 2$ this is a winning game. The required resource states can be built from Bell pairs using a combination of single-rail Bell measurements (type I fusion) and $f_{II}$. The remaining gates in order to achieve a universal gate set (a $Z_{90}$ and a
\textsc{cnot} gate) can be efficiently performed using these fusion techniques \cite{GIL05}. The resource overhead for performing gates in this way is of order 100 Bell pairs per gate.

Raussendorf and Briegel have suggested an alternative way of performing quantum computing, distinct from the usual circuit model, called cluster-state quantum computation \cite{RAU01}. It is based on measurement induced quantum evolution and so is sometimes referred to as "one-way" quantum computation. In their approach a large entangled state of a particular form, called a cluster state, is constructed first. Quantum computation is then carried out by making a series of measurements on the cluster state. For example any evolution of a single qubit can 
be simulated by: (i) preparing a string of qubits all in the states 
$\ket{\bf 0}+\ket{\bf 1}$; (ii) linking each nearest neighbour by C-S 
gates (this forms a linear cluster state), and then; (iii) measuring the single qubits in the string in sequence.   
The measurement basis chosen for each qubit depends on the single 
qubit unitaries one wishes to simulate and the result of the 
measurement of the preceding qubit. Each qubit measurement simulates 
the unitary evolution $H  Z_\theta$ where $H$ is the 
Hadamard transformation and $Z_\theta$ is a rotation about $z$. 
An arbitrary single qubit unitary can be simulated using a four qubit 
cluster state and three measurements. 

We can illustrate this by considering an arbitrary rotation about $x$, $  
X_\theta =  H  Z_\theta H$ which can be achieved 
with a 3 qubit cluster state and 2 measurements. The first 
qubit is prepared in some arbitrary state $\alpha \ket{\bf 0} + \beta \ket{\bf 1}$. A cluster state is then formed 
by applying C-S gates to the arbitrary qubit and two other qubits 
prepared in diagonal states, resulting in the state $\alpha (\ket{\bf 000}+\ket{\bf 010}+\ket{\bf 001}-\ket{\bf 011}) + \beta(\ket{\bf 100}+\ket{\bf 101}-\ket{\bf 110}+\ket{\bf 111})$. The idea is then to simulate the single 
qubit x rotation via measurement. The first 
qubit is measured in the diagonal basis: 
$\ket{D1}=\ket{\bf 0}+\ket{\bf 1}$, $\ket{D2}=
-\ket{\bf 0}+\ket{\bf 1}$. If the outcome is $D1$ then the second 
qubit is measured in the phase rotated basis: 
$\ket{R1(\theta)}=\ket{\bf 0}+\exp{i \theta} \ket{\bf 1}$, 
$\ket{R2(\theta)}=
-\ket{\bf 0}+\exp{i \theta} \ket{\bf 1}$. If, on the other hand, the outcome is $D2$ 
then the second 
qubit is measured in the phase anti-rotated basis: 
$\ket{R1(\theta)}=\ket{\bf 0}+\exp{-i \theta} \ket{\bf 1}$, 
$\ket{R2(\theta)}=
-\ket{\bf 0}+\exp{-i \theta} \ket{\bf 1}$. After these measurements 
the state of the last qubit is the same as that of the original 
qubit, but rotated about x by an angle $\theta$. However the effective 
computational basis of the qubit depends on the 
outcomes of the measurements in the following way:\\
(i) $D1$, $R1(\theta)$: the original computational basis.
(ii) $D1$, $R2(\theta)$: bit-flip of the original computational basis.
(ii) $D2$, $R1(-\theta)$: phase-flip of the original computational basis.
(ii) $D2$, $R2(-\theta)$: bit-flip and phase-flip of the original computational basis. 

By joining linear chains with CS gates to create 2-dimensional cluster states, two qubit gates can be built into the cluster, enabling universal quantum computation. The first suggestion that measurement based quantum computation could help to reduce the resources in an optical system was made by Yoran and Resnik \cite{YOR03}. Subsequently Nielsen adapted the complete cluster state approach to LOQC \cite{NIE04}. He showed that cluster states could be efficiently built up using the teleportation gates. This follows from the fact that the cluster states are able to recover from computational basis measurements in a similar (but not identical) way to that of the parity states. The application of the fusion techniques described above  \cite{BRO05} (which were in fact initially developed by Browne and Rudolph for cluster state production) further reduces the resource overhead. In this approach "mini-cluster" states are built up non-deterministically and then fused on to the main cluster in a similar way to that already described for parity states. This is perhaps the most efficient of the photonic schemes, requiring approximately 60 Bell pairs per two-qubit gate, though the exact meaning of "per gate" in the cluster state paradigm is more ambiguous than in circuit models such as the parity state approach.

In principle optical demonstrations of one-way quantum computation have now been achieved with coincidence counting. Simple cluster state computation using a 4 qubit cluster was demonstrated experimentally by Walther et al \cite{WAL05}. In this experiment the cluster state was generated directly from parametric down conversion. In other experiments the cluster states were constructed from Bell pairs by Zhang et al using the fusion technique \cite{ZHA05} and by Kiesel et al using the C-Sign gate \cite{KIE05}.

\subsubsection{Coherent States}

Finally we note that a linear optics quantum computation scheme can also be constructed using the coherent state qubits discussed in section \ref{sec:OQ}. The basic resource state required is the superposition state: $\ket{\alpha} + \ket{-\alpha}$. These, in conjunction with homodyne detection, photon counting and linear optics, are sufficient to produce a scaleable system \cite{RMM,RAL03}. As for the photonic approach the basic gates are non-deterministic and need to be scaled up by teleportation. Unlike the photonic approach the probability of success of the basic gates is much higher (80-90\%) and (as we noted in section \ref{sec:TELEPORTATION}) coherent state qubit teleportation is deterministic. This means that the over-heads for scale-up are much lower. On the other hand, the relative cost of the required superposition state resources compared to the Bell pairs needed for the photonic scheme are not known, making direct comparison's difficult. A number of groups are currently working towards demonstrations of coherent state superpositions so these issues may become clearer soon.

\subsection{Fault Tolerance}

When large scale quantum processing is considered we have to worry about the propagation of small errors inevitably introduced during gate operations. If uncorrected, such errors would grow uncontrollably and make the computation useless. The answer to this problem is fault tolerant error correction \cite{SHO95,STE96}. The idea of error correction is self-explanatory, though its implementation on quantum systems requires some care. Classically we might consider using a redundancy code such that (for example) $0 \to 0,0,0$ and $1 \to 1,1,1$. If a bit flip occurs on one of the bits we might end up with $0,1,0$ or $1,0,1$, but we can recover the original bit value by taking a majority vote. At first it may seem that such a code cannot be used for quantum mechanical systems because: (i) the no-cloning theorem \cite{WOO82} means we can not make copies of an unknown qubit and; (ii) taking the majority vote is a measurement that will collapse our quantum superposition. It turns out however that a quantum analog is possible. A quantum redundant encoding might be $\alpha \lket{0} + \beta \lket{1} \to \alpha \lket{000} + \beta \lket{111}$ where we have created an entangled state rather than copies. It is then possible, using two CNOT gates and two ancillas, to identify an error without collapsing the state, by reading out the parity of pairs of qubits. For example a bit-flip error might result in the state $\alpha \lket{001} + \beta \lket{110}$. The parity of the first two qubits will be zero whilst the parity of the second two qubits will be one, thus unambiguously identifying that an error has occured on the last qubit. Because we are measuring the parity, not the qubit value, the superposition is not collapsed. Such codes can be expanded to cope with the possibility of more than one error occurring between correction attempts and to cope with multiple types of errors. Of course the CNOT gates being used to detect and correct the errors may themselves be faulty. An error correction code is said to be {\it fault tolerant} if error propagation can be prevented even if the components used to do the error correction introduce errors themselves. Typically this is only possible if the error rate is below some level known as the {\it fault tolerant threshold}. 

The original KLM paper \cite{KNI01} showed that in principle LOQC was fault tolerant, though a general threshold was not calculated. Optical cluster state computation has also been shown to be fault tolerant, with thresholds against depolarization errors of about a hundredth of a percent \cite{DAW05}. Although such a number is daunting, the precision of optics is such that it is not inconceivable. Presently the dominant error in optical quantum processing is loss, both in components, detectors and sources. The prospects for reducing loss to such levels are remote so some effort has gone into optimizing codes specifically against loss. KLM estimated a threshold of about 1\% for loss tolerance (i.e. fault tolerance where the only error considered is loss). Remaining with the original KLM gate approach Silva et al were able to show that the loss threshold might lie as high as 11\% \cite{SIL05}. Using the parity state approach and assuming that components, sources and detectors all had an equal loss of $x \%$, Ralph et al numerically obtained a loss threshold of $x=17\%$ \cite{RAL05}. A roughly equivalent value was obtained by Varnava et al for loss tolerance of cluster states \cite{VAR05}. These nice results for loss tolerance must be treated with some caution given the tougher figures for general fault tolerance, however it is encouraging that the most resource efficient approaches also seem to display good resilience.

\section{Conclusion}

Light holds a privileged position in quantum information science as the only reasonable candidate for quantum communication. This is not just because of its mobility, but also, as we have seen,  because of the ease with which certain critical manipulations of quantum optical states can be achieved.  The scope of quantum processing tasks that can be achieved in optics has expanded rapidly in recent years leading to remarkable progress in implementing quantum
information protocols. The progress in QKD in particular
is sufficiently advanced that commercial applications are seriously considered.
Teleportation, of a quality clearly exceeding the limits set in the
absence of entanglement, has been demonstrated in both the discrete
and continuous domains. The demonstration of basic two-qubit
quantum gates is promising but is a long way short of full-scale
quantum computation. Continued advances along this path require
technical solutions to the problem of efficient single photon
production, detection and memory. 

An exciting new direction that we have not discussed much here is the possibility of hybrid optical/atomic and/or solid-state systems. It has long been recognized that optical "flying qubits" acting as a data bus can solve the connectivity problem in atomic or solid-state quantum computer architectures. Alternatively we might use the "standing qubits" as memory, whilst processing the quantum information optically. The major problem with this idea has been the interface between the standing and flying qubits. Recently significant progress has been made in this direction. For example the ion trap photon source discussed in section \ref{sec:TSPS} \cite{KEL04}, being coherent, could in principle also act as an interface. Another possibility is to use optical quantum processing to entangle distant standing qubits, thus enabling teleportation of information between distant sites or the formation of cluster states for quantum computation \cite{KOK05,DUA05}. Recent experimental progress in this direction has included the demonstration of entanglement between ions and photons \cite{BLI04}. These and other emerging technologies combined with achievements described in this review indicate a bright future for quantum information processing in optics.


\begin{thebibliography}{ABC99}
\bibitem{NOTE2} $X$, $Y$, and $Z$ are the usual Pauli operators defined in Section \ref{sec:TELEPORTATION} and an angle subscript denotes a rotation about that axis, analogous to $X_\theta$ defined in the text.
\end{thebibliography}
\end{document}